\RequirePackage{fix-cm}
\documentclass[twocolumn,epjc3]{svjour3}  
\smartqed  
\RequirePackage{graphicx}
\journalname{Eur. Phys. J. C}

\usepackage{graphicx}
\usepackage{dcolumn}
\usepackage{bm}
\usepackage{physics}
\usepackage{widetext}
\usepackage{subcaption}
\usepackage{accents}
\usepackage{amssymb}
\usepackage{xcolor}
\usepackage{soul}

\newcommand{\m}{m_{\text{P}}}

\newcommand{\Nhyp}{N_{\rm hyp}}

\newcommand{\be}{\begin{equation}} 
\newcommand{\ee}{\end{equation}}
\newcommand{\bea}{\begin{equation}\begin{aligned}} 
\newcommand{\eea}{\end{aligned}\end{equation}}
\newcommand{\ba}{\begin{eqnarray}}
\newcommand{\ea}{\end{eqnarray}}

\newcommand{\dotS}[1]{\accentset{\circ}{#1}}
\newcommand{\ddotS}[1]{\accentset{\circ\circ}{#1}}

\begin{document}

\title{Observable Gravitational Waves from Hyperkination in Palatini Gravity and Beyond
}


\author{Samuel S\'anchez L\'opez\thanksref{e1,addr1}
        \and
        Konstantinos Dimopoulos\thanksref{e2,addr1}
        \and
        Alexandros Karam\thanksref{e3,addr2}
        \and
        Eemeli Tomberg\thanksref{e4,addr1}
}

\thankstext{e1}{s.sanchezlopez@lancaster.ac.uk}
\thankstext{e2}{konst.dimopoulos@lancaster.ac.uk}
\thankstext{e3}{alexandros.karam@kbfi.ee}
\thankstext{e4}{e.tomberg@lancaster.ac.uk}


\institute{Consortium for Fundamental Physics, Physics Department, Lancaster University, Lancaster LA1 4YB, United Kingdom \label{addr1}
           \and
Laboratory of High Energy and Computational Physics, National Institute of Chemical Physics and Biophysics, R{\"a}vala pst.~10, Tallinn, 10143, Estonia \label{addr2}
}

\date{Received: 25 October 2023 / Accepted: 1 December 2023}

\maketitle

\begin{abstract}
We consider cosmology with an inflaton scalar field with an additional quartic kinetic term. Such a theory can be motivated by Palatini $R+R^2$ modified gravity. Assuming a runaway inflaton potential, we take the Universe to become dominated by the kinetic energy density of the scalar field after inflation. Initially, the leading kinetic term is quartic and we call the corresponding period hyperkination. Subsequently, the usual quadratic kinetic term takes over and we have regular kination, until reheating. We study, both analytically and numerically, the spectrum of primordial gravitational waves generated during inflation and re-entering the horizon during the subsequent eras. We demonstrate that the spectrum is flat for modes re-entering during radiation domination and hyperkination and linear in frequency for modes re-entering during kination: kinetic domination boosts the spectrum, but hyperkination truncates its peak. As a result, the effects of the kinetic period can be extended to observable frequencies without generating excessive gravitational waves, which could otherwise destabilise the process of Big Bang Nucleosynthesis. We show that there is ample parameter space for the primordial gravitational waves to be observable in the near future. If observed, the amplitude and `knee' of the spectrum will provide valuable insights into the background theory.
\keywords{Palatini gravity \and Gravitational waves \and Kination}
\end{abstract}

\section{Introduction}
\label{sec:intro}

The most compelling solution to the fine-tuning of the initial conditions of the Big Bang model is the theory of Cosmic Inflation~\cite{Starobinsky:1980te,Kazanas:1980tx,Sato:1980yn,Guth:1980zm,Linde:1981mu,Albrecht:1982wi,Linde:1983gd}. Inflation manages in a single shot to explain away the horizon and flatness problems and also to provide the primordial density perturbations necessary for the formation of the large-scale structure we observe in the Universe~\cite{Starobinsky:1979ty,Mukhanov:1981xt,Hawking:1982cz,Starobinsky:1982ee,Guth:1982ec,Bardeen:1983qw}, that is the distribution of galaxy clusters and superclusters. The primordial density perturbations reflect themselves onto the Cosmic Microwave Background (CMB) radiation, through the Sachs--Wolfe effect~\cite{Sachs:1967er}. Precision observations~\cite{Planck:2018jri,BICEP:2021xfz} of the acoustic peaks in the CMB primordial temperature anisotropy have verified in spectacular detail the predictions of Cosmic Inflation, such that the rival paradigm for structure formation (that of cosmic strings) has collapsed
\cite{Perivolaropoulos:2005wa}. Consequently, Cosmic Inflation is considered a necessary addition to the concordance model, $\Lambda$CDM, towards a standard model of cosmology.

Another generic prediction of Cosmic Inflation is coming within reach of observability in the near future. Indeed, soon after its proposal, it was realised that Cosmic Inflation gives rise to a stochastic background of primordial gravitational waves~\cite{Starobinsky:1980te,Turner:1996ck,Smith:2005mm,Boyle:2005se,Guzzetti:2016mkm,Caprini:2018mtu}. These gravitational waves (GWs) are tensor perturbations of the spacetime metric, generated in much the same way as the scalar curvature perturbations behind the primordial density perturbations, for which there is overwhelming evidence in the CMB, as mentioned above. Because of this, great interest has been developed in recent years for the observability of the inflation-produced GWs either indirectly, through the B-mode polarization of the CMB
\cite{Turner:1996ck}, or directly from interferometers
\cite{Smith:2005mm}.

Gravitational waves were predicted by Einstein’s general relativity at the beginning of the twentieth century. Almost exactly a hundred years afterwards, GWs were directly observed by LIGO (Laser Interferometer Gravitational-Wave Observatory) and Virgo in 2015~\cite{LIGOScientific:2016aoc,LIGOScientific:2016sjg}. This seminal observation heralded the birth of gravitational wave astronomy, which enables the study of compact objects, such as astrophysical black holes, which are typically shrouded by opaque  accretion disks. It also allows, in principle, a glimpse of the very early Universe, well beyond the last-scattering surface, where the CMB was emitted. As such, there is hope to detect the stochastic primordial GW background from Cosmic Inflation. Such observations will allow the study of inflation at scales much different than the ones which correspond to the CMB primordial anisotropy, opening up a new window in the understanding of fundamental physics at extremely high energies (comparable to the energy of grand unification), which is behind the process of Cosmic Inflation and remains a mystery to this day.

This has, in part, motivated a number of future GW detection missions. In the near future, Advanced LIGO (plus Virgo and KAGRA)~\cite{Harry:2010zz,VIRGO:2014yos,LIGOScientific:2014pky,LIGOScientific:2019lzm,KAGRA:2020tym} (LVK) and the space interferometer LISA (Laser Interferometer Space Antenna)~\cite{Bartolo:2016ami,Caprini:2019pxz,LISACosmologyWorkingGroup:2022jok} are coming online; the launch date of LISA is in 2037. Another space interferometer DECIGO (DECi-hertz Interferometer Gravitational wave Observatory)~\cite{Kawamura:2006up,Kawamura:2011zz,Kawamura:2020pcg} is also planned to be launched in the 2030s. More are to follow, such as BBO (Big Bang Observer)~\cite{Harry:2006fi}, a proposed successor to LISA. It seems an ideal time to investigate GW production by inflation and its potential observational signatures.

However, there is a challenge in the study of the inflation-produced primordial GW background. The background signal is too weak for any currently operational GW detector to observe, and it may be decades before an observation can be made. Indeed, were the early Universe dominated by radiation, as assumed by the concordance model, the primordial GW spectrum would be flat, \textit{i.e.} like white noise, where the GW density parameter per logarithmic frequency interval $\Omega_{\rm GW}(f)$ is constant over the range of frequencies $f$ corresponding to the GW modes that re-enter the horizon during the radiation dominated period (they have been pushed out of the horizon during inflation). The constant value of the flat spectrum is very low, and the hope of detecting in the near future such inflation-generated primordial GWs is little
\cite{Boyle:2005se}. 

Fortunately, this is not the end of our hopes for detecting primordial GWs. While there is observational evidence of the early Universe being radiation dominated, provided by the delicate process of Big Bang Nucleosynthesis (BBN) taking place a mere few seconds after the Big Bang itself, what the state of affairs was before BBN is still unknown. If the Universe's history before BBN was not dominated by radiation, then the primordial GW spectrum does not need to be flat. This opens up the possibility of a boosted GW spectrum, possible to detect even in the near future.

An early realisation of this possibility was provided by modelling quintessential inflation~\cite{Peebles:1998qn} (see Refs.~\cite{Jaman:2022bho,deHaro:2021swo} for recent reviews). Quintessential inflation aims to explain in a unified way both Cosmic Inflation in the early Universe and Dark Energy at present. Most quintessential inflation models consider non-oscillatory inflation 
\cite{Felder:1998vq,Ellis:2020krl}
driven by a scalar field (the inflaton) with a runaway potential, which can play the role of quintessence at late times and explain the accelerated expansion of the Universe at present
~\cite{Dimopoulos:2002hm,Pallis:2005hm,Cardenas:2006py,Bose:2008ew,Hossain:2014xha,WaliHossain:2014usl,Dimopoulos:2017zvq,Rubio:2017gty,Dimopoulos:2017tud,Akrami:2017cir,Geng:2017mic,Dimopoulos:2019ogl}. In such models, there is a period after the end of inflation but before reheating (\textit{i.e.} the onset of the radiation era) when the kinetic energy density of the inflaton field dominates the Universe. This period is called kination \cite{Joyce:1997fc} (see also \cite{Co:2019jts,Co:2020dya,Oikonomou:2022tux}), characterised by a stiff equation of state with a barotropic parameter \mbox{$w=p/\rho=1$}. For GW modes that re-enter the horizon during kination, the spectrum is peaked with \mbox{$\Omega_{\rm GW}(f)\propto f$} \cite{Sahni:2001qp,Giovannini:1999bh,Riazuelo:2000fc,Tashiro:2003qp,Artymowski:2017pua,Figueroa:2018twl,Oikonomou:2023bah,Oikonomou:2022ijs,Oikonomou:2023qfz}.  
Unfortunately, this peak corresponds to very high frequencies, which will be unobservable in the near future. Extending the period of kination does extend the peak to lower, possibly observable frequencies, but then the peak becomes too large and the resulting primordial GWs cannot but affect and destabilise the BBN process
\cite{Sahni:2001qp,Giovannini:1999bh,Langlois:2000ns,Dimopoulos:2002hm}.

After the direct detection of GWs, there has been much interest in considering modifications of the history of the Universe, safely before BBN, to boost the primordial GW signal at observable frequencies. In Ref. \cite{Gouttenoire:2021jhk}, it was shown that \mbox{$\Omega_{\rm GW}(f)\propto f^{-2(\frac{1-3w}{1+3w})}$}, where $w$ is the barotropic parameter of the Universe ($w=1/3$ for radiation domination). In Refs.~\cite{Co:2021lkc} and \cite{Gouttenoire:2021wzu} models were considered where there is a period of matter domination followed by kination, which would create a mountain-like peak in $\Omega_{\rm GW}$ (see also Ref.~~\cite{Gouttenoire:2021jhk}). Another possibility is to consider a stiff period after inflation that is not kination with $w=1$, but has a milder value of $w\approx 1/2$ and can be extended down to observable frequencies without destabilising the BBN because the peak is not so steep as in usual kination~\cite{Figueroa:2019paj}. A realisation of this in hybrid inflation with a non-canonical waterfall field was investigated in Refs.~\cite{Dimopoulos:2022mce,lucy}. 

In this paper, we consider a different possibility, motivated by Palatini modified gravity~\cite{Palatini1919,Ferraris1982}. The cosmological consequences of Palatini modified gravity with \mbox{${\cal L}\propto R+R^2$} and a non-minimally coupled scalar field were first considered in~\cite{Enckell:2018hmo,Antoniadis:2018ywb} in the context of inflation and subsequently in~\cite{Dimopoulos:2020pas,Dimopoulos:2022tvn,Dimopoulos:2022rdp,Antoniadis:2022cqh,Giovannini:2019mgk} in the context of quintessential inflation (see also~\cite{Tenkanen:2020dge,Gialamas:2023flv} for reviews). When switching to the Einstein frame, the scalar field obtains an additional quartic kinetic term\footnote{In Ref.~\cite{Gialamas:2022xtt} it was shown that the addition of the Holst and Holst$^2$ terms in
the usual Palatini quadratic action can generate a modification of the higher-order kinetic term.}.
In most cases considered, this term plays a negligible role in the dynamics of the scalar field. However, there are models for which this is not the case. We investigate in detail what happens when the scalar field dominating the Universe is governed by the quartic kinetic term in a period we call \emph{hyperkination}. We show that the barotropic parameter of the Universe during hyperkination is the same as that of radiation domination, $w=1/3$. As a result, in a realistic model of non-oscillatory inflation with a runaway inflaton potential, we consider a post-inflationary period of hyperkination, followed by a period of regular kination, when the kinetic energy of the inflaton is quadratic as usual. Kination is followed by radiation domination after reheating. This evolution results in a truncated peak in the GW spectrum, which can be safely extended down to observable frequencies without destabilising BBN. We calculate analytically the GW spectrum during all phases of hyperkination, kination and radiation and we verify our findings numerically. We explore the parameter space and show that we can obtain a boosted primordial GW signal with unique characteristics, which will be well-detectable by forthcoming observations. If such a signal is indeed detected, it will be a strong hint of non-canonical kinetic terms for the inflaton field from Palatini modified gravity or some other appropriate $k$-inflation or $k$-essence model.

The paper is organized as follows. In section~\ref{sec:hyperkination}, we discuss the Palatini $R^2$ models, introduce hyperkination, and embed it into the full expansion history of the Universe. In section~\ref{sec:GWs}, we consider the primordial GWs, including their initial conditions as fluctuations of the quantum vacuum. Section~\ref{sec:solution} details our analytical computation of the GW evolution. We compare our GW spectra to observational bounds in Section~\ref{sec:observations} and conclude in Section~\ref{sec:conclusions}. Throughout the paper, we use natural units with \mbox{$c=\hbar=1$} and \mbox{$8\pi G=\m^{-2}$}, where \mbox{$\m=2.43\times10^{18}\,$GeV} is the reduced Planck mass. The signature of our metric is \mbox{$(-1,+1,+1,+1)$}.

\bigskip

\section{Hyperkination}
\label{sec:hyperkination}

\subsection{\boldmath
Quartic kinetic terms from Palatini $R^2$ inflation}
\label{sec:quartic_kinetic_terms}

We begin by considering a Jordan frame action in the Palatini formulation of the form
    \begin{eqnarray} 
        S&=& \int \dd^4 x \sqrt{-g} \Big[ \frac{1}{2} h(\varphi) R + \frac{\alpha}{2} R^2 - \frac{1}{2}g^{\mu\nu}\partial_\mu\varphi\partial_\nu\varphi \nonumber \\
        &&- V(\varphi) \Big] + S_{\rm m}[g_{\mu\nu},\psi] \,,\label{eq:S_Jordan}
    \end{eqnarray}
where $\varphi$ is the inflation field and $h(\varphi)$ is its non-minimal coupling function, which usually assumes the form\footnote{Note that the non-minimal coupling $\xi$ does not affect our considerations in the following sections.} $h(\varphi) = \m^2+\xi\varphi^2$. The parameter $\alpha$ is assumed to be positive definite and we leave the potential $V(\varphi)$ unspecified. The symbol $\psi$ describes other matter components. This action was first considered in~\cite{Enckell:2018hmo,Antoniadis:2018ywb} in the context of inflation and then in~\cite{Dimopoulos:2020pas,Dimopoulos:2022tvn,Dimopoulos:2022rdp,Antoniadis:2022cqh} in the context of quintessential inflation (see also~\cite{Tenkanen:2020dge,Gialamas:2023flv} for reviews).

In the Palatini formulation of gravity, the connection $\Gamma$ and the metric $g_{\mu\nu}$ are independent variables. The Ricci tensor $R_{\mu\nu} (\Gamma)$ only depends on the connection and the Ricci scalar is defined as $R \equiv g^{\mu\nu} R_{\mu\nu}(\Gamma)$. The connection $\Gamma$ can be determined by varying the action~\eqref{eq:S_Jordan} but, due to the non-minimal coupling function $h(\varphi)$ and the $\alpha R^2$ term, it will differ from the standard Levi-Civita form.

Following~\cite{Enckell:2018hmo,Antoniadis:2018ywb}, we can eliminate the $\alpha R^2$ term by introducing an auxiliary scalar field $\chi \equiv 2\alpha R$. Then, by performing a Weyl transformation of the form $\bar{g}_{\mu\nu} = \Omega^2 g_{\mu\nu} = [ \chi + h(\varphi) ] g_{\mu\nu}$ we bring the action to the canonical form with a minimally coupled scalar field. The resulting action will depend on two scalar fields: $\varphi$ and $\chi$. However, in contrast to the usual metric formalism, the auxiliary field $\chi$ is non-dynamical in the Palatini formalism. This means that one can vary the action with respect to $\chi$, solve the resulting constraint equation, and then eliminate $\chi$ altogether from the action. 

After this procedure, the resulting action in the Einstein frame reads~\cite{Enckell:2018hmo,Antoniadis:2018ywb}
    \begin{eqnarray} 
        S&=&\int \dd^4 x \sqrt{-\bar{g}}\Big[\frac{\m^2}{2}\bar{R}-\frac{1}{2}(\bar{\partial}\phi)^2 + \frac{\alpha}{4}\frac{h^2+4\alpha V}{h^2\m^4}(\bar{\partial}\phi)^4 \nonumber \\
        &&-U\Big]+S_{\text{m}}[\Omega^{-2}\bar{g}_{\mu\nu},\psi] \, ,\label{eq:actionnoncanonicaleinstein}
    \end{eqnarray}
where
\begin{equation} \label{eq:V_Einstein}
     U \equiv \frac{V\m^4}{h^2+4\alpha V}\,,
\end{equation}
and we employed a field redefinition of the form
    \begin{equation} \label{eq:field_transform}
        \frac{\dd \phi}{\dd \varphi}=\sqrt{\frac{h(\varphi)\m^2}{h(\varphi)^2+4\alpha V(\varphi)}} \, 
    \end{equation}
in order to render the quadratic kinetic term canonical, where the bars indicate quantities in the Einstein frame.
Note that the process of transforming from the Jordan to the Einstein frame has generated a quartic kinetic term\footnote{Note that, in the context of Palatini gravity, models that contain a non-minimal derivative coupling term $G_{\mu\nu} \partial^\mu \varphi \partial^\nu \varphi$~\cite{Gialamas:2020vto} or $R_{(\mu\nu)}R^{(\mu\nu)}$ terms~\cite{Annala:2021zdt,Gialamas:2021enw} in the Jordan frame, can lead to actions similar to~\eqref{eq:actionnoncanonicaleinstein} in the Einstein frame after applying a disformal transformation of the metric.} and a modified potential $U$ which will in general display a plateau for growing $V$, approaching the asymptotic value $\m^4/(4\alpha)$~\cite{Enckell:2018hmo}. Also, importantly, in the present work, we concentrate on the early era when the other matter components $\psi$ are a perfect fluid of radiation. In this limit, the coupling between the inflaton and the matter action in the last term of Eq. \eqref{eq:actionnoncanonicaleinstein} disappears~\cite{Dimopoulos:2022rdp}.

Neglecting the last term for the moment, we can rewrite the action as
    \begin{equation}
    \label{eq:action_kessence}
        S=\int \dd^4 x \sqrt{-\bar{g}}\qty[\frac{\m^2}{2}\bar{R} + P\left( \phi, X \right)] \,,
    \end{equation}
with
    \begin{equation}
        P\left( \phi, X \right) = X + L(\phi) X^2 - U \,,
    \end{equation}
where 
\begin{equation}
X \equiv - \frac{(\bar{\partial}\phi)^2}{2} \qquad \text{and} \qquad L(\phi) \equiv \frac{\alpha}{4}\frac{h^2+4\alpha V}{h^2\m^4}\,.
\end{equation}
The action in Eq.~\eqref{eq:action_kessence} belongs to the general class of $k$-inflation \cite{Armendariz-Picon:1999hyi} (where inflation is kinetically driven) or $k$-essence~\cite{Chiba:1999ka,Armendariz-Picon:2000nqq,Armendariz-Picon:2000ulo} (where the non-canonical kinetic terms can behave as quintessence).\footnote{In Ref.~\cite{Gialamas:2019nly}
it was shown that the Palatini $R^2$ models share common features with $k$-inflation models.}

Varying the action in Eq. \eqref{eq:actionnoncanonicaleinstein} we can obtain the equation of motion for $\phi$, which reads~\cite{Enckell:2018hmo}
\begin{widetext}
    \begin{eqnarray} 
        \Big[1+3\alpha\qty(1+\frac{4\alpha V}{h^2})\frac{\dot{\phi}^2}{\m^4}\Big]\ddot{\phi} + 3\Big[1+\alpha\qty(1+\frac{4\alpha V}{h^2})\frac{\dot{\phi}^2}{\m^4}\Big] \bar{H} \dot{\phi}+ 3\alpha^2\frac{\dot{\phi}^4}{\m^4}\frac{\dd}{\dd\phi}\Big(\frac{V}{h^2}\Big) + \frac{\dd}{\dd\phi}U  = 0 \, .\label{eq:phi_eom_Einstein}
    \end{eqnarray}
\end{widetext}
Then, from the non-zero components of the energy-momentum tensor we can obtain the energy density and pressure of the field, which read~\cite{Karam:2021sno}
    \begin{equation} 
    \label{eq:rho_p_V}
    \begin{aligned}
        \bar{\rho}_\phi &= \frac{1}{2}\qty[1+\frac{3}{2}\alpha\qty(1+\frac{4\alpha V}{h^2})\frac{\dot{\phi}^2}{\m^4}]\dot{\phi}^2 + U \, , \\
        \bar{p}_\phi &= \frac{1}{2}\qty[1+\frac{1}{2}\alpha\qty(1+\frac{4\alpha V}{h^2})\frac{\dot{\phi}^2}{\m^4}]\dot{\phi}^2 - U \, .
    \end{aligned}
    \end{equation}
To complete the equations of motion, the Hubble parameter can be written as 
    \begin{equation} \label{eq:H}
        3\m^2 \bar{H}^2 = \bar{\rho}_\phi \,.
    \end{equation}
Again, the above equations differ from those of a standard canonical scalar field due to the higher-order kinetic terms. In the limit $\alpha \to 0$ they reduce to the minimal case. The bars are dropped in what follows to avoid clutter. Unless otherwise stated we always work in the Einstein frame.

The plateau in $U$ mentioned above is ideal for slow-roll inflation, and can easily produce CMB observables compatible with observations for simple forms of the potential $V$ \cite{Enckell:2018hmo,Antoniadis:2018ywb}. However, it restricts the inflationary---and thus post-inflationary---energy density to values lower than $\m^4/(4\alpha)$. Unfortunately, this severely restricts the parameter space considered in the following sections. One way to overcome this problem is to consider an $\alpha$ that experiences a drastic change at the end of inflation but remains constant afterwards. This is possible if $\alpha$ depends on a degree of freedom that changes its value when inflation ends. A toy model discussing this possibility is presented in \ref{appx:toy}. Another example of a model describing the full inflationary history may be the one studied in Ref. \cite{Dimopoulos:2022tvn}, as long as it is enhanded with the hybrid mechanism discused in \ref{appx:toy}. Moreover, we point out that the Palatini $R^2$ models considered here act as an inspiration for the extra quartic kinetic terms in the action, but our analysis is more general, and we do not specify the details of the inflationary part of the model.

\subsection{Kinetic domination}
\label{sec:kinetic_domination}
While the quartic kinetic terms in Eq.~\eqref{eq:actionnoncanonicaleinstein} are negligible during slow-roll inflation~\cite{Enckell:2018hmo}, they may play an important role in the post-inflationary Universe. We consider next such a scenario; a period of kinetic domination, where the potential $V$ is negligible and the field rolls forward freely. In this limit, Eqs.~\eqref{eq:phi_eom_Einstein} and \eqref{eq:H} become
\begin{eqnarray}
    \Big(1+3\alpha\frac{\dot{\phi}^2}{\m^4}\Big)\ddot{\phi} &+& 3\Big(1+\alpha\frac{\dot{\phi}^2}{\m^4}\Big) H \dot{\phi} = 0 \, ,\nonumber \\
    3H^2\m^2 &=& \rho_\phi \, ,     \label{eq:phi_eom_kination}
\end{eqnarray}
and
\begin{eqnarray}
    \rho_\phi &=& \frac{1}{2}\Big(1+\frac{3}{2}\alpha\frac{\dot{\phi}^2}{\m^4}\Big)\dot{\phi}^2 \, , \nonumber \\
    p_\phi &=& \frac{1}{2}\Big(1+\frac{1}{2}\alpha\frac{\dot{\phi}^2}{\m^4}\Big)\dot{\phi}^2 \, .     \label{eq:rho_p_Einstein}
\end{eqnarray}
It is instructive to change the time variable to the number of elapsing e-folds $N = \ln a$, with $\dd N = H \dd t$, and eliminate $H$. We can assume $\dot{\phi} > 0$ without loss of generality. The field time derivatives are related as\footnote{A prime denotes a derivative with respect to $N$ in this section only. In the rest of the paper, it denotes a derivative with respect to the conformal time $\eta$, $\dd \eta = \dd t/a$. As an exception, $\phi_0'$ in Eq. \eqref{eq:hyperkination_approximate_solution}, which is used throughout the paper, is always equal to $\phi_0'=\dot{\phi}/H$ evaluated at the end of inflation.}
\begin{equation} \label{eq:phi_dot_relation}
    \dot{\phi} = \m^2\sqrt{\frac{2(6\m^2-\phi'^2)}{3\alpha\phi'^2}} \, .
\end{equation}
Note that, due to the scaling with the heavily $\dot{\phi}$-dependent $H$, the limit $\dot{\phi} \to 0$ corresponds to $\phi' \to \sqrt{6}\m$, and $\dot{\phi} \to \infty$ corresponds to $\phi' \to 0$. Eqs.~\eqref{eq:phi_eom_kination} and \eqref{eq:rho_p_Einstein} become
\begin{eqnarray} 
    \phi'' &&= \frac{\phi'(6\m^2 - \phi'^2)(12\m^2 + \phi'^2)}{6\m^2(12\m^2 - \phi'^2)} \, , \nonumber \\
    \rho_{\phi} &&= \frac{2\m^6}{\alpha \phi'^2}\left(\frac{6\m^2}{\phi'^2}-1\right)\, , \nonumber \\
    p_{\phi} &&= \frac{2\m^6}{3\alpha\phi'^2}\left(\frac{6\m^2}{\phi'^2}+1\right) - \frac{2\m^4}{9\alpha} \, , \nonumber \\
    w_{\phi} &&= \frac{1}{9}\left(3 + \frac{\phi'^2}{\m^2}\right) \, ,\label{eq:N_eom}
\end{eqnarray}
where $w_{\phi} \equiv p_{\phi}/\rho_{\phi}$ is the barotropic parameter of the field. Note that $\alpha$ dropped out of the equation of motion: changing $\alpha$ rescales the time and energy density but leaves quantities like $\phi$, $N$, and $w_{\phi}$ untouched.

\begin{figure*}[h]
     \centering
         \includegraphics[width=0.6\textwidth]{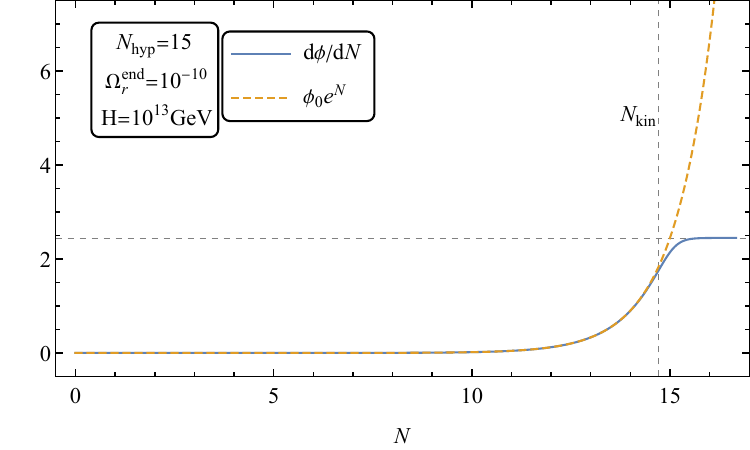}
         \caption{$N$-derivative of the field obtained from the numerical simulation (full blue line) and its initial approximation given in Eq. \eqref{eq:hyperkination_approximate_solution} (dashed orange line) as functions of $N$. The dashed vertical line, labelled $N_{\rm kin}$, corresponds to the time at which kination starts in the numerical simulation, defined here as the moment at which both addends inside the parenthesis in the energy density in Eq. \eqref{eq:rho_p_Einstein} become equal, while the dashed horizontal line corresponds to $\phi'=\sqrt{6}\m$. In the legend, $H$ denotes the Hubble parameter at the end of inflation $H_{\rm end}$.}
         \label{fig:dphi_w_example}
\end{figure*}

If $\dot{\phi}$ is small---that is,  $\frac{3}{2}\alpha\dot{\phi^2} \ll \m^4$ and $\phi' \approx \sqrt{6}\m$---the quartic extra kinetic terms are small, and Eq.~\eqref{eq:N_eom} give
\begin{eqnarray} 
    &&\phi'' \approx 6(\sqrt{6}\m - \phi') \quad \Rightarrow \quad \phi' \approx \sqrt{6}\m\left(1 - ce^{-6N}\right) \, , \nonumber \\
    &&\rho_\phi \propto (6\m^2 - \phi'^2) \propto e^{-6N} \propto a^{-6} \, , \quad
    w_\phi \approx 1 \, ,\label{eq:kination_dynamics}
\end{eqnarray}
where $c$ is an integration constant and we are concerned with the large $N$ limit. We see that $\phi' = \sqrt{6}\m$ is an attractor. It corresponds to standard \emph{kination} \cite{Spokoiny:1993kt,Joyce:1996cp,Joyce:1997fc,Pallis:2005hm,Pallis:2005bb,Gomez:2008js} with a quickly diluting energy density and $w_\phi \approx 1$.

In the opposite limit of $\frac{3}{2}\alpha\dot{\phi^2} \gg \m^4$ and $\phi'\approx 0$, the quartic kinetic terms dominate, and Eq.~\eqref{eq:N_eom} gives
\begin{eqnarray}
    &&\phi'' \approx \phi' \quad \Rightarrow \quad \phi' \approx ce^{N} \propto a \, , \nonumber \\
    &&\rho_\phi \propto (\phi')^{-4} \propto a^{-4} \, , \quad
    w_\phi \approx \frac{1}{3} \, . \label{eq:hyperkination_dynamics}
\end{eqnarray}
We name this phase \emph{hyperkination}. The extra kinetic terms modify the dynamics so that the energy density dilutes only as fast as radiation with $w_\phi \approx 1/3$. 

Hyperkination only lasts for a limited time. As spatial expansion dilutes the field's kinetically dominated energy density, $\dot{\phi}$ decreases and $\phi'$ grows. The quartic kinetic terms are diluted faster than the quadratic ones, and eventually the latter take over. Consequently, the field transitions into standard kination. We can use Eqs.~\eqref{eq:kination_dynamics} and \eqref{eq:hyperkination_dynamics} to approximate the time evolution of $\phi'$ as it approaches the kination attractor as
\begin{equation} \label{eq:hyperkination_approximate_solution}
    \phi' \approx
    \begin{cases}
        \phi'_{0} e^N & N < \ln(\sqrt{6}/\phi'_{0}) \, , \\
        \sqrt{6}\m & N > \ln(\sqrt{6}/\phi'_{0}) \, ,
    \end{cases}
\end{equation}
where $\phi'_{\text{0}}$ is the initial value of $\phi'$ at $N=0$, taken below to be the end of inflation. Tuning $\phi'_{\text{0}}$ lets us modify the length of hyperkination, which we define as\footnote{With the restriction $\rho_\phi < \m^4/(4\alpha)$ discussed at the end of section~\ref{sec:quartic_kinetic_terms}, we would have $\phi'_0 > 2\m$ at $N=0$, and Eq. \eqref{eq:hyperkination_approximate_solution} restricts hyperkination to last less than $0.20$ e-folds, a negligible amount. As mentioned, we omit this restriction in this paper.} 
\begin{equation}
N_{\rm hyp} \equiv \ln(\sqrt{6}\m/\phi'_{0})\,.
\label{Nhyp}
\end{equation}
Figure~\ref{fig:dphi_w_example} compares Eq.~\eqref{eq:hyperkination_approximate_solution} to a numerical solution of Eq.~\eqref{eq:N_eom} in an example case.

Due to the exponential growth of $\phi'$, the transition  from hyperkination to kination is fast. Let us define the beginning of standard kination $N_{\rm kin}$ as the moment when both addends inside the parenthesis in the energy density in Eq. \eqref{eq:rho_p_Einstein} become equal. Using Eqs.~\eqref{eq:phi_dot_relation} and \eqref{eq:hyperkination_approximate_solution}, this condition reads
\begin{equation} \label{eq:Nhyp_vs_Nkin}
    1=\frac{3\alpha \dot{\phi}^2}{2\m^4}=e^{2(N_{\rm hyp}-N_{\rm kin})}-1\, \Leftrightarrow \, N_{\rm kin}=N_{\rm hyp}-\ln{\sqrt{2}} \, .
\end{equation}
Thus, $N_{\rm hyp}\simeq N_{\rm kin}$ and we conclude that it is a good approximation to assume an instantaneous transition between hyperkination and kination.

We end this section with a relation between $\alpha$, the energy density at the start of hyperkination (end of inflation) $\rho_{\rm end}$, and $N_{\rm hyp}$. Equation \eqref{eq:N_eom} together with the definition of $N_{\rm hyp}$ gives
\begin{equation} \label{eq:rho_end_vs_Nhyp}
    \alpha \rho_{\rm end}=\frac{\m^4(1-e^{-2N_{\rm hyp}}) }{3 e^{-4N_{\rm hyp}}}\simeq \frac{\m^4 e^{4 N_{\rm hyp}}}{3} \, ,
\end{equation} 
where we have assumed a non-negligible duration for hyperkination $\mathcal{O}(N_{\rm hyp})\sim 1$ in the last step. Note that $N_{\rm hyp}=0$ corresponds to $\alpha=0$, as it should.
\begin{figure*}
     \centering
     \begin{subfigure}[b]{0.49\textwidth}
         \centering
         \includegraphics[width=\textwidth]{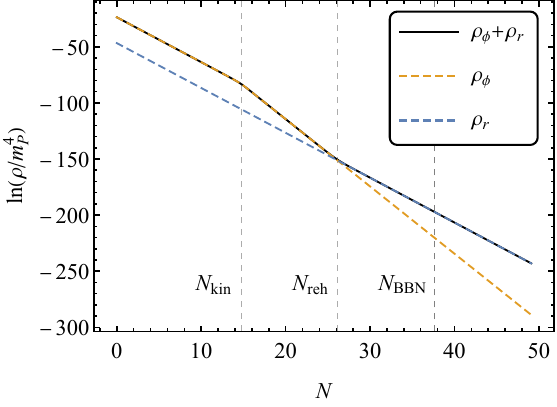}
     \end{subfigure}
     \begin{subfigure}[b]{0.49\textwidth}
         \centering
         \includegraphics[width=\textwidth]{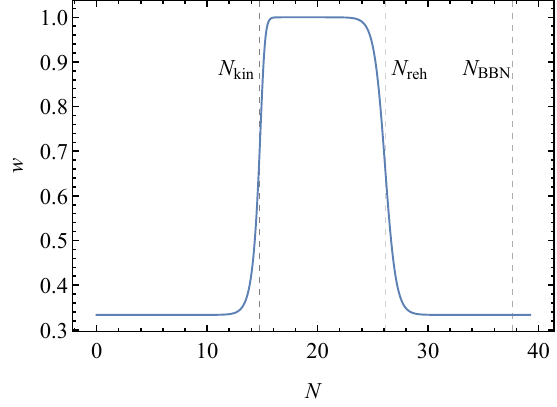}
     \end{subfigure}
     \caption{Left:  Logarithm of the energy density of the Universe (full black), the field (dashed orange) and the background radiation fluid (dashed blue) as a function of the number of e-folds calculated from the end of inflation, obtained by numerically solving the system. Right: Barotropic parameter of the Universe from the same computation. The vertical dashed lines correspond to the start of kination, reheating, and the BBN. The parameters for both panels are $N_{\rm hyp}=15$, $\Omega_{\rm r}^{\rm end}=10^{-10}$ and $H=10^{13}$ GeV.}
    \label{fig:w_and_rho}
\end{figure*}

\subsection{Full cosmic evolution}
\label{sec:cosmic_evolution}
Let us now embed a period of kination into a full history of the Universe. Initially, during cosmic inflation, the field energy density is dominated by potential energy. Once inflation ends, the potential drops to zero and the field's velocity increases as the potential energy is transformed into kinetic energy. In typical models, the field is trapped into a potential minimum, oscillating there and decaying into a thermal bath of particles, reheating the Universe. In our models of interest, the post-inflationary potential is of the runaway type---that is, flat and low---and the field keeps rolling onward under kinetic domination. If the quartic kinetic terms dominate, this phase starts with hyperkination, transitioning into standard kination later, as described above.

To reheat the Universe, we assume a small amount of radiation is produced at the end of inflation e.g. through Ricci reheating \cite{Dimopoulos:2018wfg,Opferkuch:2019zbd,Bettoni:2021zhq}. During hyperkination, the radiation energy density is diluted as fast as that of the field, $\rho_{\rm{r},\phi} \propto a^{-4}$, so radiation stays subdominant. However, when standard kination starts, the field energy density dilutes faster, $\rho_\phi \propto a^{-6}$, and the radiation fraction grows until it overtakes the field. The Universe is reheated and radiation domination starts. We assume this to take place at high energies, above the BBN temperature $T_\text{BBN} \approx 1 \ \rm{MeV}$; afterwards, the Universe follows the standard $\Lambda$CDM expansion history.

The behaviour of the system can be solved from the Friedmann equations
\begin{equation} \label{eq:Friedmann_phi_and_rad}
    3H^2\m^2 = \rho_{\rm{r}} + \rho_\phi \, , \quad \rho_{\rm{r}} = 3(H_{\rm end})^2\m^2 \times \Omega_{\rm r}^{\rm end}\left(\frac{a}{a_{\rm end}}\right)^{-4} \, ,
\end{equation}
combined with the first equations from Eqs.~\eqref{eq:phi_eom_kination} and \eqref{eq:rho_p_Einstein}.
Here $\Omega_{\rm r}^{\rm end}$ is the radiation energy density fraction (parameter) at the end of inflation and $a_{\rm end}$ and $H_{\rm end}$ are the scale factor and Hubble parameter at the end of inflation. Figure \ref{fig:w_and_rho} shows the behaviour of the energy densities solved numerically; for details on the numerical implementation, see \ref{sec:numerics}. It also shows the corresponding evolution of the barotropic parameter $w$, defined as the ratio between the total pressure and energy density of the Universe, taking values from $w=1/3$ (hyperkination) to $w=1$ (standard kination) back to $w=1/3$ (radiation domination).

In summary, we assume a cosmological evolution where inflation is followed by two phases: hyperkination and kination, in this order. Reheating, which takes places at temperatures larger than $T_{\rm BBN}$, signals the end of these phases. After reheating, the conventional cosmic evolution with radiation and matter dominated eras follows\footnote{Some authors (see Ref. \cite{Co:2021lkc} and the discussion on page 6 in Ref. \cite{Oikonomou:2023bli}) have considered that the stiff era takes place after BBN, but before recombination. This would relax the lower bound for the temperature of the stiff phase to $T<6\,\text{keV}$. However, this possibility is not considered in the present work.}.

The non-standard expansion history opens the door for new phenomenology. For one, it changes the matching between scales in the early and late Universe. Indeed, when inflation is followed by a stiff cosmological era with barotropic parameter $w$, the number of inflationary e-folds is increased by \cite{Dimopoulos:2017zvq,Oikonomou:2022tux,Dimopoulos:2020pas,Dimopoulos:2019gpz}
\begin{equation}
    \Delta N=\frac{3w-1}{3(1+w)}\ln{\left(\frac{V_{\rm end}^{1/4}}{T_{\rm reh}}\right)}.
\end{equation}
It follows that hyperkination, for which $w=1/3$, has a vanishing contribution. This is not the case for kination, with $w=1$.
Thus, in our scenario we have
\begin{equation}
    \Delta N=\frac13\ln{\left(\frac{\rho_{\rm kin}^{1/4}}{T_{\rm reh}}\right)},
    \label{DN}
\end{equation}
where $\rho_{\rm kin}\ll V_{\rm end}$ is the energy density at the end of hyperkination and the onset of kination proper. Typically, this increases the remaining number of inflationary efolds after the
cosmological scales exit the horizon to at most \mbox{$N_*\simeq 65$}, which implies \mbox{$\Delta N\lesssim 5$}, something that must be taken into account when calculating the inflationary observables.\footnote{Such an increase has some effect on the inflationary observables, but this effect is minimal. For example, in Starobinsky inflation \cite{Starobinsky:1980te} or Higgs inflation \cite{Bezrukov:2007ep} (or $\alpha$-attractors \cite{Kallosh:2013yoa}), the scalar spectral index is \mbox{$n_s\simeq 1-\frac{2}{N_*}$}. With \mbox{$N_*=60$} this results in \mbox{$n_s=0.966$}.
If we have \mbox{$N_*=65$} instead, then \mbox{$n_s=0.969$}, which is still within the 1-$\sigma$ contour of the Planck satellite observations \cite{Akrami:2018odb}.}

All in all, the CMB scales exit the Hubble radius approximately 60--65 e-folds before the end of inflation instead of the standard 50--60, see e.g.~\cite{Dimopoulos:2020pas,Liddle:2003as}. This affects inflationary model building, although the effects are mitigated with respect to the standard kination scenario.
In addition, the spectrum of primordial GWs is altered in ways that are sensitive to the duration of hyperkination.

\section{Gravitational waves}
\label{sec:GWs}

\subsection{Tensor perturbations}
\label{sec:tensor_perturbations}
To study the behaviour of GWs, we write the metric tensor as $g_{\mu\nu}=a^2\left(\eta_{\mu\nu}+h_{\mu\nu}\right)$, where $\eta_{\mu\nu}$ is the Minkowski metric so that $a^2\eta_{\mu\nu} \equiv \bar{g}_{\mu\nu}$ is the unperturbed FLRW metric, and $h_{\mu\nu}$ is a small perturbation. We expand the action in Eq.~\eqref{eq:actionnoncanonicaleinstein} to second order in $h_{\mu\nu}$, keeping only the tensor modes\footnote{The tensor perturbations obey $\partial_{\mu}h^{\mu\nu}=0$ and $h^{\mu}_{\phantom{\mu}\mu}=0$.}, which evolve independently of other perturbations in linear perturbation theory. The result is (compare to, e.g., \cite{Riotto:2002yw})
\begin{eqnarray}
    \label{eq:S2_basic}
    &&\delta^{(2)}S
    = -\frac{\m^2}{8}\int \text{d}^4x\sqrt{-\bar{g}}\bar{g}^{\alpha\beta}\partial_{\alpha}h^{\mu\nu}\partial_{\beta}h_{\mu\nu} \\
    \label{eq:S2_polars}
    &&=- \!\!\! \sum_{s=\oplus,\otimes}\frac{\m^2}{4}\int \text{d}^4x\sqrt{-\bar{g}} \bar{g}^{\alpha\beta}\partial_{\alpha}h^{s}\partial_{\beta}h^{s}\\
    \label{eq:S2_h}
    \label{eq:S2_Fourier}
    &&= \sum_{s=\oplus,\otimes} \frac{\m^2}{4}\int \dd \eta \, a^2 \int \dd^3k \left(|h^s_{\vec{k}}{}'|^2 - k^2|h^s_{\vec{k}}|^2\right) \, ,
\end{eqnarray}
where $\eta$ is the conformal time related to the cosmic time $t$ by $\dd t = a \, \dd \eta$, and a prime denotes a derivative with respect to $\eta$. Here $s$ indexes the two gravitational wave polarisations, and the polarization amplitudes $h^s$ are defined through the Fourier decompositions
\begin{equation} \label{eq:hs_Fourier}
    h^s(\vec{x}) = \int \frac{\dd^3 k}{(2\pi)^{3/2}} \, h^s_{\vec{k}} \, e^{i\vec{k}\cdot\vec{x}} \, ,
\end{equation}
so that $h^s_{\vec{k}}$ describes oscillations of a given polarization in directions perpendicular to the wave vector $\vec{k}$.

The amplitudes $h^s$ behave as massless scalar fields, up to normalization, following the Klein--Gordon equation
\begin{equation} \label{eq:klein_gordon}
    {h^s}'' + 2\mathcal{H}{h^s}' + \nabla^2 h^s = 0
\end{equation}
with wave solutions. Here $\mathcal{H} \equiv a'/a$ and $\nabla^2 \equiv \partial_i\partial_i$ where $i$ is summed over the spatial indices. The corresponding energy-momentum tensor is
\begin{eqnarray} 
     &&T^\mathrm{GW}_{\mu\nu}= -\frac{2}{\sqrt{-\bar{g}}}\frac{\delta (\delta^{(2)} S)}{\delta \bar{g}^{\mu\nu}} \nonumber \\
    &&= \sum_{s=\oplus,\otimes} \frac{\m^2}{2}\left( \partial_\mu h^s \partial_\nu h^s - \frac{1}{2}\bar{g}_{\mu\nu}\bar{g}^{\alpha\beta}\partial_\alpha h^s \partial_\beta h^s \right) \, ,\label{eq:GW_energy_momentum}
\end{eqnarray}
so that the GW energy density reads
\begin{equation} \label{eq:GW_energy_density}
    \rho_\mathrm{GW}
    = a^{-2}T^\mathrm{GW}_{00}
    = \sum_{s=\oplus,\otimes} \frac{\m^2}{4a^2} \left[(h^s{}')^2 + (\nabla h^s)^2\right] \, .
\end{equation}

\subsection{Quantization}
\label{sec:quantization}
The primordial GWs originate as quantum vacuum fluctuations during inflation. To quantize them, we first go to the canonically normalized variables $v^s=\m a h^s/\sqrt{2}$, so that (after integration by parts) the action in Eq.~\eqref{eq:S2_polars} becomes
\begin{eqnarray}
    \label{eq:S2_v}
    &&\delta^{(2)}S= \nonumber \\
    &&=\sum_{s=\oplus,\otimes}\frac{1}{2}\int \text{d}^3x\text{d}\eta \left[- \eta^{\alpha\beta}\partial_{\alpha}v^{s}\partial_{\beta}v^{s}+\frac{a''}{a}(v^s)^2\right] \\
    \label{eq:S2_v_Fourier}
    &&= \sum_{s=\oplus,\otimes} \frac{1}{2}\int \dd \eta \, \dd^3k \left[|v^s_{\vec{k}}{}'|^2 - \left(k^2 - \frac{a''}{a}\right)|v^s_{\vec{k}}|^2\right] \, .
\end{eqnarray}
This is the Minkowski space action for a free field with mass $a''/a$, quantized the standard way by writing
\begin{equation} \label{eq:v_hat}
    \hat{v}^s(\eta,\vec{x})=\int \frac{\text{d}^3k}{(2\pi)^{3/2}}\left[v_k^s(\eta)\hat{a}^s_{\vec{k}}e^{i\vec{k}\cdot\vec{x}}+v_k^{s*}(\eta)\hat{a}^{s^{\dagger}}_{\vec{k}}e^{-i\vec{k}\cdot\vec{x}}\right],
\end{equation}
where $\hat{a}^s_{\vec{k}}$, $\hat{a}^{s^{\dagger}}_{\vec{k}}$ are the ladder operators following the canonical commutation relations
\begin{equation} \label{eq:commutation_relations}[\hat{a}^{s'}_{\vec{k}'},\hat{a}^{s^{\dagger}}_{\vec{k}}]=\delta^{s's}\delta^{(3)}(\vec{k}'-\vec{k}) \, .
\end{equation}
Time evolution is delegated to the mode functions $v_k^s$, which follow the Mukhanov--Sasaki equations derived from Eq.~\eqref{eq:S2_v_Fourier},
\begin{equation} \label{eq:v_eom}
    v^s_k{}'' + \left(k^2 - \frac{a''}{a} \right)v^s_k = 0 \, .
\end{equation}
Note that, due to the ladder operators, the mode functions $v^s_k$ differ in normalization from the classical Fourier modes $v^s_{\vec{k}}$. Abusing the notation slightly, we differentiate these by writing $k$ instead of $\vec{k}$ as the mode function index---in an FLRW background, the quantum mode functions only depend on the magnitude of the wave vector and not its direction. Analogously, we define $\hat{h}^s = \sqrt{2}\hat{v}^s/(a\m$), $h^s_k = \sqrt{2}v^s_k/(a\m)$.

Deep inside the Hubble radius, $k \gg \mathcal{H}$, the GWs do not feel the expansion of space, the mass term $a''/a$ is negligible, and Eq.~\eqref{eq:v_eom} has the standard vacuum solution
\begin{equation} \label{eq:Bunch-Davies}
    v_k^s = \frac{1}{\sqrt{2k}}e^{-i k \eta} \, , \qquad v_k^s{}' = -ikv_k^s \, .
\end{equation}
When the mode functions follow Eq.~\eqref{eq:Bunch-Davies}, the state annihilated by $\hat{a}_k^s$ is the Bunch--Davies vacuum \cite{Birrell:1982ix}; we take the perturbations to start in this vacuum state during inflation. Over their cosmic evolution, the modes stretch and exit the Hubble radius, evolving beyond Eq.~\eqref{eq:Bunch-Davies}. After inflation, they re-enter the Hubble radius, this time following the general sub-Hubble form
\begin{equation} \label{eq:sub_Hubble_modes_general}
    v^s_k=\frac{1}{\sqrt{2k}}\left[\lambda_{+}(k)e^{-ik\eta}+\lambda_{-}(k)e^{ik\eta}\right] \, .
\end{equation}
We will solve the coefficients $\lambda_\pm(k)$ for a given cosmic history in section \ref{sec:mode_functions}; since the Mukhanov--Sasaki equation conserves the Wronskian of its solutions, we have $|\lambda_+|^2 - |\lambda_-|^2 = 1$, set by the initial vacuum in Eq.~\eqref{eq:Bunch-Davies}. The coefficient $\lambda_-$ contains the GW excitations, the part beyond the vacuum solution in Eq.~\eqref{eq:Bunch-Davies}.

Let us next consider the energy density of the GWs induced by the above process. The late-time GW energy density is dominated by high-$k$, sub-Hubble modes, for which Eq.~\eqref{eq:sub_Hubble_modes_general} applies. Using this result, we replace $h^s$ by $\hat{h}^s$ in the energy-momentum tensor in Eq.~\eqref{eq:GW_energy_momentum} and compute its expectation value. The result is
\begin{eqnarray}
    \expval{\hat{\rho}_\mathrm{GW}}
    &&= \sum_{s=\oplus,\otimes} \frac{\m^2}{2} \int \frac{(\dd \ln k) \, k^3}{4\pi^2a^2} \qty(|h'^s_k|^2 + k^2|h^s_k|^2) \nonumber\\
    &&\approx \!\!\int\displaylimits_{k=\mathcal{H}} \frac{(\text{d}\ln{k})}{2\pi^2}\frac{k^4}{a^4}\left(\abs{\lambda_{+}}^2+\abs{\lambda_{-}}^2\right) \nonumber \\
    &&= \!\!\int\displaylimits_{k=\mathcal{H}} \frac{(\text{d}\ln{k})}{\pi^2}\frac{k^4}{a^4}\left(\abs{\lambda_{-}}^2+\frac{1}{2}\right) \, , \label{eq:rho_GW}
\end{eqnarray}
where we used the Wrosnkian condition, and the fact that the integration limit $k>\mathcal{H}$ restricts us to sub-Hubble modes. In the last line, we have taken the polarization sum (starting from the Bunch--Davies vacuum, $\lambda_\pm$ are identical for both polarizations).

Note that, regardless of $\lambda_\pm$, the final term of $1/2$ makes Eq.~\eqref{eq:rho_GW} diverge for large $k$---this is the usual energy density vacuum divergence of quantum field theory. One can regularize the result by normal ordering the ladder operators in $\hat{\rho}_\mathrm{GW}$. However, this has to be done with the late-time ladder operators which annihilate the late-time Bunch--Davies vacuum. These are related to the original ladder operators $\hat{a}_k^s$ by a Bogoliubov transformation; for a detailed discussion, see e.g. Ref.~\cite{Birrell:1982ix}. The regularized energy density becomes
\begin{equation} \label{eq:rho_GW_regularized}
    \expval{\hat{\rho}_\mathrm{GW}}
    \approx \!\!\int\displaylimits_{k=\mathcal{H}} \frac{(\text{d}\ln{k})}{\pi^2}\frac{k^4}{a^4} \abs{\lambda_{-}}^2 \, .
\end{equation}
In practice, all of our modes of interest are highly excited with $|\lambda_-| \gg 1$, so that Eqs.~\eqref{eq:rho_GW} and \eqref{eq:rho_GW_regularized} are approximately equal. In this limit, the vacuum contribution is negligible and the GWs are essentially classical.

\subsection{Energy density scaling and the problem with kination}

From Eq.~\eqref{eq:rho_GW_regularized}, we see that the sub-Hubble GWs scale as radiation, with $\rho_{\mathrm{GW}} \propto a^{-4}$, as expected for massless degrees of freedom. In cosmology with a standard expansion history, only a small amount of GWs are generated during inflation, and they always stay subdominant compared to the background radiation energy density. However, during kination, the background dilutes faster than radiation, and the gravitational wave fraction grows. The resulting gravitational wave spectrum is peaked and tends to either clash with bounds on the number of relativistic degrees of freedom during BBN or be hard to observe in gravitational wave experiments \cite{Sahni:2001qp,Giovannini:1999bh,Riazuelo:2000fc,Tashiro:2003qp,Artymowski:2017pua,Figueroa:2018twl}. In the following sections, we will demonstrate that adding a period of hyperkination helps with this issue, opening a wider parameter space for allowed GW spectra.

\section{Analytical solution}
\label{sec:solution}

\subsection{Solving the background}
\label{sec:background_solution}

Let us move on to solve the GW spectrum analytically. The first step is to solve the background dynamics, in particular the scale factor $a$, in the presence of radiation, as a function of the conformal time $\eta$. This provides us with $a''/a$, allowing us to later solve the Mukhanov--Sasaki equation for the GW mode functions.

The scale factor evolves through different epochs during the cosmic history: inflation, hyperkination, kination, and radiation domination. The transitions between the epochs, assumed to be instantaneous, happen at conformal times $\eta_{\rm end}$ (end of inflation and start of hyperkination), $\eta_{\rm kin}$ (end of hyperkination and start of kination), and $\eta_{\rm reh}$ (end of kination and start of radiation domination, \textit{i.e.}, reheating), which we will also solve in terms of the model parameters below. We use the same indices to refer to various variables evaluated at these times. We  require the continuity of $a(\eta)$ and its derivative at the transition times; between them, we solve $a(\eta)$ from
\begin{equation} \label{eq:a_eta_def}
    \dd \eta = \frac{\dd t}{a} = \frac{\dd a}{a^2H} = \frac{\dd a}{a^2}\sqrt{\frac{3\m^2}{\rho}} \, .
\end{equation}
If we know how the Universe's energy density $\rho$ scales in $a$, we can integrate and invert Eq.~\eqref{eq:a_eta_def} to obtain $a(\eta)$ epoch by epoch. We will normalize the scale factor so that 
 \begin{equation} \label{eq:scalefactornormalization}
     a(\eta_{\rm end})=1 \, ,
\end{equation}
and write $a=e^N$, so that $N$ counts the e-folds since the end of inflation.

For inflation, we assume a generic slow-roll inflationary phase, with the end of inflation $\eta_{\rm end}<0$ determined by the usual condition
\begin{equation} \label{eq:epsilon}
    \epsilon \equiv - \frac{\dot{H}}{H^2} = 1 \, ,
\end{equation}
where $\epsilon$ is the first Hubble slow-roll parameter. For the reader's benefit, we will express our GW mode functions as a function of $\epsilon$, approximated to be constant. In the example spectra we consider in section~\ref{sec:observations}, we work in the pure de Sitter limit $\epsilon=0$. To avoid clutter (and slightly abusing the notation), we will use $H$, as for pure de Sitter, to refer to the Hubble parameter at the end of inflation $H_{\rm end}$.

For hyperkination, we get $\rho(N)$ from Eq.~\eqref{eq:N_eom}, where $\phi'$ follows the first branch of Eq.~\eqref{eq:hyperkination_approximate_solution} and we write the initial field velocity $\phi_0'$ in terms of $N_{\rm hyp}$ as explained below the equation. For kination and radiation domination, we use the standard results $\rho \propto a^{-6}$ and $\rho \propto a^{-4}$. With these, the full behaviour of the scale factor becomes

\begin{widetext}
\begin{equation} \label{eq:a}
    a = 
    \begin{cases}
        \left[-\frac{1}{(1-\epsilon)H\eta}\right]^{1/(1-\epsilon)} \, , & \eta \leq \eta_{\rm end} \, , \\
        e^{N_{\rm hyp}}\sin\left[e^{-N_{\rm hyp}}(H\eta +1) + \sin^{-1}e^{-N_{\rm hyp}}\right] \, , & \eta_{\rm end} \leq \eta \leq \eta_{\rm kin} \, , \\
        a_\text{kin}\sqrt{2\mathcal{H}_\text{kin}(\eta-\eta_\text{kin})+1} \, , & \eta_{\rm kin} \leq \eta \leq \eta_{\rm reh} \, , \\
        a_{\text{reh}}[\mathcal{H}_{\text{reh}}(\eta-\eta_{\text{reh}})+1] \, , & \eta_{\rm reh} \leq \eta \, .
    \end{cases}
\end{equation}
\end{widetext}

For the hyperkination expression, we used
\begin{equation}
    \eta_{\rm end}=-\frac{1}{(1-\epsilon)H}\simeq -\frac{1}{H} \, ,
\end{equation}
which follows from Eq.~\eqref{eq:scalefactornormalization} and the first line in Eq.~\eqref{eq:a}. We also used Eq.~\eqref{eq:rho_end_vs_Nhyp} with $3H^2\m^2 = \rho_{\rm end}$ to eliminate $\alpha$. For a long hyperkination period with $N_{\rm hyp} \gtrsim 1$, we can approximate the expression as
\begin{equation} \label{eq:a_hyperkination_approx}
    a(\eta) \simeq e^{N_{\rm hyp}}\sin\left[e^{-N_{\rm hyp}}(H\eta +2)\right]\simeq H\eta +2 \, ,
\end{equation}
where the right-hand-side is exactly the scale factor for a radiation-dominated universe, compare to the last line in Eq.~\eqref{eq:a}. Note that the last approximation stops being valid at large times $\eta\sim e^{N_{\rm hyp}}/H$ and one needs to use the middle expression instead. This is the case below, when we obtain an analytical estimate for $\eta_{\rm kin}$.

For kination and radiation domination, the constants in Eq.~\eqref{eq:a} are to be read off from the end values during the previous phase. Using Eq.~\eqref{eq:a_hyperkination_approx}, we have
\begin{eqnarray}
    a_\text{kin} &&= e^{N_{\rm hyp}}\sin\left[e^{-N_{\rm hyp}}(H\eta_{\rm kin} +2)\right] \, , \nonumber \\
    a_\text{reh} &&= a_\text{kin}\sqrt{2\mathcal{H}_{\text{kin}}(\eta_{\rm reh}-\eta_{\rm kin})   + 1} \, , \label{eq:kination_radation_constants_a}
\end{eqnarray}
and
\begin{eqnarray}
    \mathcal{H}_{\text{kin}} &&= \frac{He^{-N_{\rm hyp}}}{\tan\left[e^{-N_{\rm hyp}}(H\eta_{\rm kin} +2)\right]} \, , \nonumber \\
    \mathcal{H}_{\text{reh}} &&= \frac{\mathcal{H}_{\text{kin}}}{2\mathcal{H}_{\rm kin}(\eta_{\rm reh} - \eta_{\rm kin}) + 1} \, .     \label{eq:kination_radation_constants_H}
\end{eqnarray}

In practice, it is a good approximation to use
\begin{equation}
    a_{\rm kin}=H\eta_{\rm kin}+2 \, , \qquad \mathcal{H}_{\text{kin}} = \frac{H}{H\eta_\text{kin} + 2} \, .
\end{equation}

Let us next estimate the conformal times for the rest of the transition points. We do this by solving an equation where $a$ is expressed in two different ways, through Eq.~\eqref{eq:a} and through a condition related to our model parameters.

As a reminder, we define the beginning of kination as the time at which both addends inside the parenthesis in the energy density in Eq.~\eqref{eq:rho_p_Einstein} become equal. Since this happens at large times $\eta\sim e^{N_{\rm hyp}}/H$, we use the middle expression in Eq.~\eqref{eq:a_hyperkination_approx} together with Eq.~\eqref{eq:Nhyp_vs_Nkin} to obtain
\begin{equation}
    \label{eq:solving_eta_kin}
    a_{\rm kin}=e^{N_{\rm hyp}}\sin\left[e^{-N_{\rm hyp}}(H\eta_{\rm kin} +2)\right]=e^{N_{\rm kin}}=\frac{e^{N_{\rm hyp}}}{\sqrt{2}},
\end{equation}
so that
\begin{equation}
    \label{eq:eta_kin}
    \eta_{\rm kin}=\frac{\frac{\pi}{4}e^{N_{\rm hyp}}-2}{H}\simeq\frac{\pi e^{\Nhyp}}{4H} \, .
\end{equation}

The time of reheating $\eta_{\rm reh}$ can be estimated by noting that the total energy density during kination scales as $\rho\propto a^{-6}$, while that of the radiation scales as $\rho_{\rm r}\propto a^{-4}$. Thus, the density parameter of radiation during kination scales as $\Omega_{\rm r}\propto a^{2}$. By reheating, radiation is the dominant component, that is,
\begin{equation}
    \label{eq:solving_eta_reh}
    1
    \approx\Omega_{\rm r}^{\rm reh}
    \approx\Omega_{\rm r}^{\rm kin}\left(\frac{a_{\rm reh}}{a_{\rm kin}}\right)^2
    = \Omega_{\rm r}^{\rm end}\frac{2H^2\eta_{\rm kin}\eta_{\rm reh}}{e^{2N_{\rm hyp}}/2}\, ,
\end{equation}
so that
\begin{equation}
    \label{eq:eta_reh}
    \eta_{\rm reh}=\frac{e^{\Nhyp}}{\pi \Omega_{\rm r}^{\rm end}H} \, ,
\end{equation}
where we used $\Omega_{\rm r}^{\rm kin}\approx \Omega_{\rm r}^{\rm end}$, since the field and radiation redshift similarly during hyperkination, together with the approximation $|\eta_{\rm end}| \ll \eta_{\rm kin}\ll \eta_{\rm reh}$ yielding $a_{\rm reh}\approx H\sqrt{2\eta_{\rm kin}\eta_{\rm reh}}$ from Eqs.~\eqref{eq:kination_radation_constants_a} and \eqref{eq:kination_radation_constants_H}. We also used Eq.~\eqref{eq:solving_eta_kin} for $a_{\rm kin}$ and Eq.~\eqref{eq:eta_kin} for $\eta_{\rm kin}$.

\subsection{The gravitational wave mode functions}
\label{sec:mode_functions}

\begin{figure*}[h]
     \centering
    \begin{subfigure}[b]{0.45\textwidth}
         \centering
         \includegraphics[width=\textwidth]{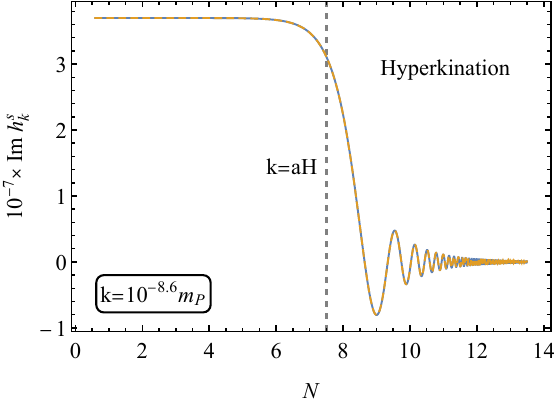}
     \end{subfigure}\hfill\hfill
     \begin{subfigure}[b]{0.45\textwidth}
         \centering
         \includegraphics[width=\textwidth]{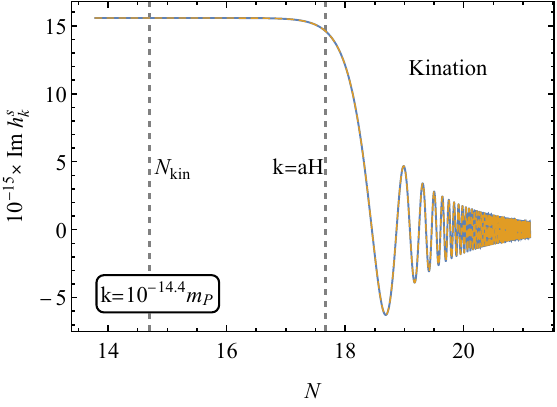}
     \end{subfigure}
     \begin{subfigure}[b]{0.45\textwidth}
         \centering
         \includegraphics[width=\textwidth]{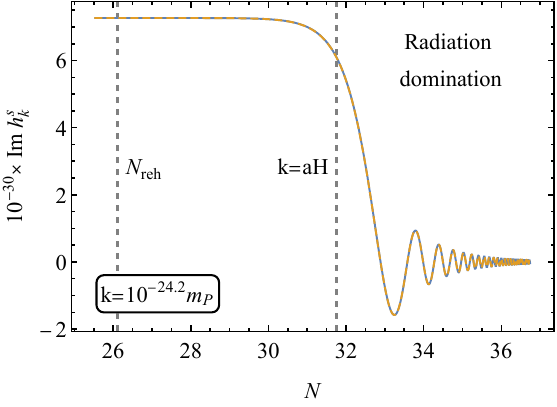}
     \end{subfigure}\hfill\hfill
     \begin{subfigure}[b]{0.45\textwidth}
         \centering
         \includegraphics[width=\textwidth]{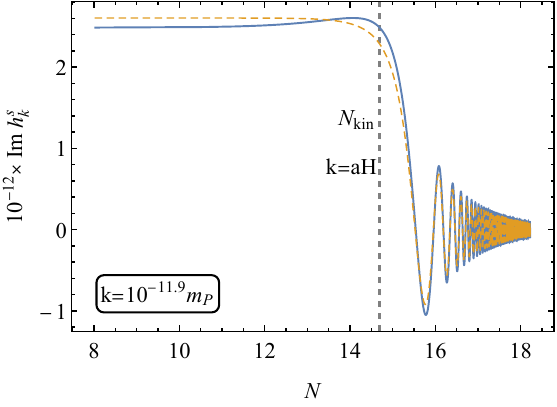}
     \end{subfigure}
     \caption{Comparison between the analytical solution (solid blue lines) and its numerical counterpart (dashed orange) of the imaginary part of the mode functions $h_k^s$ as a function of the elapsing number of e-folds when the mode enters the horizon during the hyperkination (top left), kination (top right) and radiation domination (bottom left) periods. The match is excellent, except when the wavenumber of the mode is comparable to the horizon size at a transition (bottom right). The vertical dashed lines represent the time of horizon crossing $k=aH$ and the times at which kination starts $N_{\rm kin}$ and reheating happens $N_{\rm reh}$. The parameters for all panels are $N_{\rm hyp}=15$, $\Omega_{\rm r}^{\rm end}=10^{-10}$ and $H=10^{13}$ GeV.}
      \label{fig:modefunctionscompared}
\end{figure*}

The next step is to obtain expressions for the GW mode functions. We proceed by matching the solutions and their derivatives at the transitions between epochs. To simplify the expressions, we do the matching in the super-Hubble limit, which gives an excellent approximation except for modes entering the horizon around the transitions. Our goal is to obtain the coefficients $\lambda_-(k)$ from Eq.~\eqref{eq:sub_Hubble_modes_general} for each mode so that we can read off their asymptotic, sub-Hubble behaviour. We report the details of the somewhat technical calculations in \ref{appendixA}, while in the present section we simply give the main results, as well as a comparison between the analytical and numerical solutions in Fig. \ref{fig:modefunctionscompared} (for details on the numerics see \ref{sec:numerics}).

We can summarize the scale factor time dependence from the last section as
\begin{equation}
    a=\left(\frac{\eta}{\eta_c}\right)^{1/2-\nu} \, ,
    \qquad
    \nu \equiv \frac{3(w-1)}{2(1+3w)} \, ,
\end{equation}
where $w$ is the corresponding barotropic parameter of the Universe, so that $\nu=3/2$ ($w=-1$) for de Sitter, $\nu = 3/2+\epsilon \equiv \nu_I$ for a more realistic quasi-de Sitter inflation \cite{Caprini:2018mtu},  $\nu=0$ ($w=1$) for kination and $\nu=-1/2$ ($w=1/3$) for hyperkination and radiation domination. We then get
\begin{equation} \label{eq:scalefactor2}
    \frac{a''}{a}=-\left(\frac{1}{4}-\nu^2\right)\frac{1}{\eta^2} \, .
\end{equation}
The constants $\eta_c$ can be read from the previous section, giving
\begin{equation} \label{eq:a_prime_prime}
    \frac{a''}{a} =
    \begin{cases}
        \frac{2+3\epsilon}{\eta^2} \, , & \eta \leq \eta_{\rm end} , \\
        0 \, , & \eta_{\rm end} \leq \eta \leq \eta_{\rm kin} , \\
        -\frac{1}{4z^2} \, , & \eta_{\rm kin} \leq \eta \leq \eta_{\rm reh} , \\
        0 \, , & \eta_{\rm reh} \leq \eta ,
    \end{cases}
\end{equation}
where we defined for kination
\vspace{0.6cm}
\begin{equation} \label{eq:z}
    z\equiv \eta-\frac{\eta_{\rm kin}}{2}+\frac{1}{H} \, .
\end{equation}
\vspace{0.1cm}

Note that $a''=0$ during hyperkination. This feature is shared with the period of radiation domination, during which the spectrum is flat, a result that was originally derived in Ref. \cite{Sahni:1990tx}. Therefore we expect the peak from kination to be truncated by a secondary plateau.

With this, we can proceed to solve the Mukhanov--Sasaki equation \eqref{eq:v_eom}. Making the change of variables $x=k\eta$ ($x=-k\eta$ during inflation when $\eta < 0$) and redefining the mode functions as $v=\sqrt{x}g$, it can be recast as a Bessel equation
\vspace{0.6cm}
\begin{equation} \label{asiphdbfajsdf}
    x^2\frac{\text{d}^2g}{\text{d}x^2}+x\frac{\text{d}g}{\text{d}x}+(x^2-\nu^2)g=0 \, ,
\end{equation}
\vspace{0.1cm}

\noindent the most general solution of which is given by
\vspace{0.6cm}
\begin{equation}
    g(x)=c_1 H_{\nu}^{(1)}(x)+c_2H_{\nu}^{(2)}(x) \, ,
\end{equation}
\vspace{0.1cm}

\noindent where $H_{\nu}^{(1)}$ and $H_{\nu}^{(2)}$ are Hankel functions of the first and second kind respectively. Using the values of $\nu$ from above, the solutions during inflation, hyperkination, kination, and radiation domination become
\begin{widetext}
\vspace{0.4cm}
\begin{equation} \label{eq:v_k_cases}
v_k^s(\eta) =
\begin{cases}
    \sqrt{\frac{\pi}{4k}}\sqrt{-k\eta}e^{i\frac{\pi}{4}\left(1+2\nu_I \right)}H_{\nu_I}^{(1)}(-k\eta) \, , & \eta\leq\eta_{\rm end} \, , \\ 
    \frac{1}{\sqrt{2k}}\left[\alpha_{+}(k)e^{-ik\eta}+\alpha_{-}(k)e^{ik\eta}\right] \, , & \eta_{\rm end}\leq\eta\leq\eta_{\rm kin} \, , \\
    \sqrt{\frac{\pi z}{4}}\left[\beta_{+}(k)e^{-i\pi/4}H_{0}^{(2)}(kz) + \beta_{-}(k)e^{i\pi/4}H_0^{(1)}(kz)\right] \, , &  \eta_{\rm kin}\leq\eta\leq\eta_{\rm reh} \, , \\
    \frac{1}{\sqrt{2k}}\left[\gamma_{+}(k)e^{-ik\eta}+\gamma_{-}(k)e^{ik\eta}\right]\, , & \eta_{\rm reh}\leq\eta \, ,
\end{cases}
\end{equation}
\vspace{0.4cm}
\end{widetext}
where we fixed the coefficients $c_{1,2}$ during inflation so that in the initial sub-Hubble regime, $-k\eta \gg 1$, the mode functions obey the Bunch--Davies vacuum conditions in Eq.~\eqref{eq:Bunch-Davies}. The constants and phases in the other branches have been chosen so that the coefficients $\alpha_\pm$, $\beta_\pm$, and $\gamma_\pm$ correspond to the $\lambda_{\pm}$ of Eq.~\eqref{eq:sub_Hubble_modes_general} in the late sub-Hubble limit $k\eta\gg 1$. Their values are fixed by requiring the continuity of $v_k^s$ and its derivative at the transition times $\eta_{\rm end}$, $\eta_{\rm kin}$, and $\eta_{\rm reh}$. Matching the branches in the super-Hubble limit yields
\begin{eqnarray}
    \alpha_{\pm}(k)&=&\mp \frac{f(\epsilon)}{2}\left(\frac{H}{k}\right)^{2+\epsilon} \, ,   \label{ad9fausdfaa} \\
    \beta_{\pm}(k)&=&2i e^{\pm i\pi/4}\alpha_{-}(k)\sqrt{\frac{k\eta_{\rm kin}}{\pi}} \, , \\
    \gamma_{\pm}(k)&=&\mp \alpha_{-}(k)\sqrt{\frac{\eta_{\rm kin}}{2z_{\rm reh}}} \, ,     \label{a9dufbauisdf}
\end{eqnarray}
where
\begin{equation} \label{eq:f}
    f(\epsilon)\equiv e^{i\pi\epsilon/2}\frac{\Gamma(3/2+\epsilon)}{\Gamma(3/2)}2^{\epsilon} \, ,
\end{equation}
and $z_{\rm reh}\simeq \eta_{\rm reh}$ is $z$ from \eqref{eq:z} evaluated at $\eta_{\rm reh}$.
For the scale-invariant case with $\epsilon\to 0$, $f(\epsilon)\to 1$, the moduli squared of the coefficients take the simplified forms
\begin{eqnarray} 
    \abs{\alpha_{-}(k)}^2&=&\frac{H^4}{4k^4} \, , \nonumber\\
    \abs{\beta_{-}(k)}^2&=&\frac{H^4}{\pi k^4}k\eta_{\rm kin} \, ,\nonumber\\
    \abs{\gamma_{-}(k)}^2&=&\frac{H^4}{4k^4}\frac{\eta_{\rm kin}}{2\eta_{\rm reh}} \, . \label{eq:alpha_beta_gamma_de_Sitter}
\end{eqnarray}

Note that since we did the matchings at the super-Hubble limit, the expressions in Eqs.~\eqref{ad9fausdfaa}--\eqref{a9dufbauisdf} and \eqref{eq:alpha_beta_gamma_de_Sitter} only apply for modes that are super-Hubble during the corresponding transition. To find the final behaviour of a mode, we take the last transition where this applies, track the following mode function from Eq.~\eqref{eq:v_k_cases} to the sub-Hubble limit, where it takes the form in Eq.~\eqref{eq:sub_Hubble_modes_general}, and equate the $\alpha_-$, $\beta_-$, or $\gamma_-$ with the coefficient $\lambda_-$. Indeed, after a mode has settled to its asymptotic sub-Hubble behaviour, its evolution is trivial---redshfiting gently like radiation---and it won't be sensitive to further changes in the equation of state of the Universe.

From the Mukhanov--Sasaki solutions in Eq.~\eqref{eq:v_k_cases} we can also deduce the metric perturbations $h_k^s$.
Using the scale factor expressions, $a\simeq H\eta$, $a\simeq H\sqrt{2\eta_{\rm kin}\eta}$, and $a\simeq H\sqrt{\eta_{\rm kin}/(2\eta_{\rm reh})}\eta$ during hyperkination, kination, and radiation domination, respectively, and using Eqs.~\eqref{ad9fausdfaa}--\eqref{a9dufbauisdf}, we get
\begin{equation} \label{eq:h_k_cases}
h_k^s(\eta) = 
\begin{cases}
    \frac{i H}{\m k^{3/2}}f(\epsilon)\left(\frac{k}{H}\right)^{-\epsilon}j_{0}(k\eta) \, , & \eta_{\rm end}\leq\eta\leq\eta_{\rm kin} \, , \\
    \frac{i H}{\m k^{3/2}}f(\epsilon)\left(\frac{k}{H}\right)^{-\epsilon}J_0(kz) \, , & \eta_{\rm kin}\leq\eta\leq\eta_{\rm reh} \, ,\\
    \frac{iH}{\m k^{3/2}}f(\epsilon)\left(\frac{k}{H}\right)^{-\epsilon}j_{0}(k\eta) \, , & \eta_{\rm reh}\leq\eta \, ,
\end{cases}
\end{equation}
where $j_{0}(k\eta)=\sqrt{\pi/(2k\eta)}J_{1/2}(k\eta)=\sin{k\eta}/(k\eta)$ is a spherical Bessel function of the first kind and $J_0$ is a Bessel function of the first kind. For a comparison with the numerical solutions in the scale-invariant case, see Fig. \ref{fig:modefunctionscompared}. We do not include the inflationary metric perturbations in Eq.~\eqref{eq:h_k_cases} as they do not simplify as nicely as the others.

Note that in the super-Hubble limit, all the expressions in Eq.~\eqref{eq:h_k_cases} freeze to
\begin{equation} \label{adufubauisdf}
    h_k^s(\eta)
    \, \xrightarrow{k|\eta| \to 0} \,
    \frac{i H}{\m k^{3/2}}f(\epsilon)\left(\frac{k}{aH}\right)^{-\epsilon} 
    \, \xrightarrow{\epsilon \to 0} \,
    \frac{i H}{\m k^{3/2}} \, ,
\end{equation}
where the last one is the standard scale-invariant result. Note that this result holds also for inflation. In principle, one can use this as an initial condition and solve the Klein--Gordon equation \eqref{eq:klein_gordon} to obtain Eq.~\eqref{eq:h_k_cases} separately in each phase without the matching procedure described above\footnote{In particular, GWs at the CMB scales stay frozen throughout the kination and hyperkination periods and are thus not affected by the non-standard background evolution. The same is true for the curvature perturbation $\mathcal{R}$---see \cite{Garriga:1999vw} for a linear treatment of $\mathcal{R}$ in a model with a non-standard kinetic sector, Appendix~B of \cite{Karam:2021sno} for an application to Palatini $R^2$ models, and \cite{Lyth:2004gb} for a general proof that $\mathcal{R}$ freezes at super-horizon scales.}. One can then use Eq.~\eqref{eq:rho_GW} to obtain the unregularized GW energy density. We use this method in our numerical solutions. The expressions for $\alpha_\pm$, $\beta_\pm$, and $\gamma_\pm$ are still needed to regularize the integral in Eq.~\eqref{eq:rho_GW}, and they are the conventional way to express the GW excitations in the literature.

\section{Gravitational wave observations}
\label{sec:observations}

\subsection{Gravitational wave spectrum}
\label{sec:GW_spectrum}

We are finally in a position to calculate the spectral energy density of the primordial GW background. It is defined as
\begin{equation} \label{eq:Omega_GW}
    \Omega_{\rm GW}(k,\eta) \equiv \frac{1}{\rho(\eta)}\frac{\text{d} \rho_{\rm GW}(k, \eta)}{\text{d}\ln{k}} = \frac{1}{\rho(\eta)}\frac{k^4\abs{\lambda_{-}(k)}^2}{\pi^2a^4(\eta)}  \, ,
\end{equation}
where $\rho$ is the total energy density of the Universe and $\rho_{\rm GW}(k, \eta)$ is the contribution to the GW energy density from modes around $k$, given by Eq.~\eqref{eq:rho_GW_regularized} for the dominant, sub-Hubble modes. Here $\lambda_-$ is to be matched to $\alpha_-$, $\beta_-$, or $\gamma_-$ as explained above.

To evaluate Eq.~\eqref{eq:Omega_GW} at a specific time, we note that the radiation energy density can be written as (remember our normalization $a_{\rm end} = 1$)\footnote{We neglect the change in the effective number of relativistic species contributing to the entropy $g_{*S}(T)$ and to the energy density $g_{*}(T)$. This introduces an additional mild scale dependence into the spectrum. For further details, we refer the reader to Ref. \cite{Watanabe:2006qe}, and in particular to Fig. 4 therein.}
\begin{equation}
    \label{eq:deriving_rho_a4_eq}
    \rho_{\rm r}(\eta) = \Omega_{\rm r}(\eta) \rho(\eta) = \Omega_{\rm r}^{\rm end}\rho_{\rm end} a^{-4}(\eta)\, , \\
\end{equation}
so that
\begin{equation}
    \label{eq:rho_a4}
    \rho(\eta)a^{4}(\eta) = \rho_{\rm end} \frac{\Omega_{\rm r}^{\rm end}}{\Omega_{\rm r}(\eta)} \, .
\end{equation}
In particular, using the current radiation temperature and total energy density, $T_0=2.7~\text{K}=0.23\times 10^{-9}~\text{MeV}$ and $\rho_0=1.05\times 10^{-120}\m^4$ \cite{Aghanim:2018eyx}, we obtain  $\rho_{\rm r}^0=8.79\times 10^{-125}\m^4$ and $\Omega_{\rm r}^0=8.37\times 10^{-5}$. We use the index `0' to refer to quantities today. With this and the de Sitter limit results in Eq.~\eqref{eq:alpha_beta_gamma_de_Sitter} together with Eqs.~\eqref{eq:eta_kin} and \eqref{eq:eta_reh}, the GW spectrum today becomes
\begin{widetext}
\begin{equation} \label{eq:Omega_GW_spectrum_today}
\Omega_{\rm GW}(k, \eta_0) = 
\begin{cases}
    \frac{\Omega_{\rm r}^0}{96}\left(\frac{H}{\m}\right)^2, & k<k_{\rm reh} , \\
    \frac{\Omega_{\rm r}^0}{12\pi^2\Omega_{\rm r}^{\rm end}}\left(\frac{H}{\m}\right)^2\frac{k}{H}e^{N_{\rm hyp}} , & k_{\rm reh} < k < k_{\rm kin} , \\
    \frac{\Omega_{\rm r}^0}{12\pi^2 \Omega_{\rm r}^{\rm end}}\left(\frac{H}{\m}\right)^2 , & k_{\rm kin}<k<k_{\rm end} .
\end{cases}
\end{equation}
\end{widetext}
Below, we will refer to the different branches as $\Omega_{\rm GW}^{\rm rad}$, $\Omega_{\rm GW}^{\rm kin}$, and $\Omega_{\rm GW}^{\rm hyp}$. The boundary values are given by $k=\mathcal{H}$ at the end of inflation and at the transition times. Using Eqs.~\eqref{eq:kination_radation_constants_a} and \eqref{eq:kination_radation_constants_H}, we get
\begin{eqnarray}
    k_{\rm end}&=&H \, ,\nonumber \\
    k_{\rm kin}&\simeq&\frac{1}{\eta_{\rm kin}}=\frac{4H}{\pi e^{N_{\rm hyp}}} \, , \nonumber \\
    k_{\rm reh}&\simeq&\frac{1}{2\eta_{\rm reh}}=\frac{\pi\Omega_{\rm r}^{\rm end}H}{2e^{N_{\rm hyp}}} \, , \label{eq:transition_ks}
\end{eqnarray}
where we approximated $\abs{\eta_{\rm end}} \ll \eta_{\rm kin} \ll \eta_{\rm reh}$.

In our figures, we show the spectrum as a function of $f$, the GW frequency today. To relate $f$ to our wavenumber\footnotemark $k$, we use Eq.~\eqref{eq:deriving_rho_a4_eq} and $\rho=3H^2\m^2$, yielding
\begin{equation} \label{eq:f_vs_k}
    f=\frac{k}{2\pi a_0 }=\frac{1}{2\pi} \left(\frac{\Omega_{\rm r}^0H_0^2}{\Omega_{\rm r}^{\rm end}H^2}\right)^{1/4}k \, .
\end{equation}
An important frequency is the one that corresponds to BBN. It does not depend on the early expansion history, and we can solve it explicitly as
\begin{eqnarray} \label{adoijfbaibdfwef}
    f_{\rm BBN}=&&\frac{1}{2\pi}\frac{a_{\rm BBN}H_{\rm BBN}}{a_0}=\frac{1}{2\pi}\left(\frac{\rho_{\rm r}^0}{\rho_{\rm BBN}}\right)^{1/4}\left(\frac{\rho_{\rm BBN}}{3\m^2}\right)^{1/2}\nonumber \\
    \simeq&& 1.36\times 10^{-11}\,\text{Hz} \, ,\label{eq:fbbn}
\end{eqnarray}
where we used $\rho_{\rm BBN}\simeq 3\times 10^{-86}\m^4$. We present $f_{\rm BBN}$ as a vertical dotted line in our graphs. 

\begin{figure*}
\centering
\includegraphics[width=0.6\textwidth]{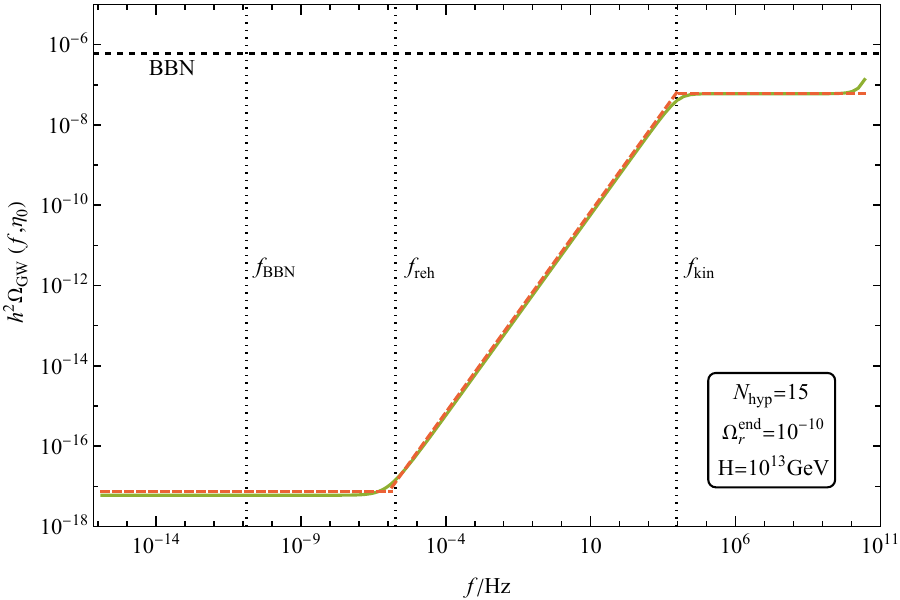}
\caption{Analytical spectral energy density of the primordial GWs (dashed orange) and its numerical counterpart (full green). For details on the numerical solution, we refer the reader to \ref{sec:numerics}. The vertical dotted lines represent the frequencies associated with the start of kination, reheating and BBN, while the horizontal dashed line represents the BBN bound on the spectrum. The numerical spectral energy density is not well resolved at the largest frequencies because the modes re-entering the horizon right after inflation are never frozen as assumed in the code. This leads to the unphysical upslope around $10^{11}\,$Hz. The parameters used are $N_{\rm hyp}=15$, $\Omega_{\rm r}^{\rm end}=10^{-10}$ and $H=10^{13}$ GeV.} 
\label{fig:spectrum}
\end{figure*}

We show a comparison between the numerical and analytical spectra, for an example set of parameters, in Fig. \ref{fig:spectrum}. We see that the analytical expressions for the spectrum are very accurate. In Fig. \ref{jdvasdbfiasdf} we present some example analytical spectra superimposed with the sensitivity curves for future GW experiments. 

From Eqs.~\eqref{eq:Omega_GW_spectrum_today} and \eqref{eq:transition_ks}, we can straightforwardly understand the shape of the spectrum. The height of the first plateau, corresponding to hyperkination, is given by the combination $H^2/(\Omega_{\rm r}^{\rm end}\m^2)$, \textit{i.e.}, the larger the energy density at the end of inflation and the smaller the reheating efficiency, the larger the energy density spectrum amplitude will be. The third free parameter of our theory, the number of e-folds of hyperkination $N_{\rm hyp}$, controls the length of the boosted spectrum; the longer the hyperkination period lasts, the more stretched the boosted spectrum is. In contrast, the height of the second plateau depends on $H^2/\m^2$, \textit{i.e.}, it depends on the energy scale at the end of inflation only, the standard result from a scenario with no period of kinetic domination, originally derived in Ref. \cite{Sahni:1990tx}. Both plateaus are connected via a region growing linearly with the frequency $f$, corresponding to the kination period. At large frequencies, the spectrum is cut off at the last mode to be excited by inflation. At small frequencies, there is no cutoff; the first line in Eq.~\eqref{eq:Omega_GW_spectrum_today} applies to all modes that re-enter during radiation domination.

Although it is easier to understand the shape of the spectrum in terms of $N_{\rm hyp}$, the free parameter in the action in Eq.~\eqref{eq:actionnoncanonicaleinstein} is $\alpha$. For this reason, we present below our results regarding the parameter space of the theory in terms of $\alpha$ and not $N_{\rm hyp}$. The two are related by Eq.~\eqref{eq:rho_end_vs_Nhyp}. For completeness, we present here the spectrum in terms of $\alpha$, and with $k$ replaced with $f$:
\begin{widetext}
\begin{equation} \label{aipsdbfbiuadsf}
\Omega_{\rm GW}(f,\eta_0) = 
\begin{cases}
    \frac{\Omega_{\rm r}^0}{96}\left(\frac{H}{\m}\right)^2 \, , & f<f_{\rm reh} \, , \\
    \left(\frac{\Omega_{\rm r}^{0}}{\Omega_{\rm r}^{\rm end}}\right)^{3/4}\frac{H^{3/2}}{6\pi H_0^{1/2}\m^2}\left(\frac{1+\sqrt{1+36\alpha H^2/\m^2}}{2}\right)^{\frac{1}{2}}f \, , & f_{\rm reh}<f<f_{\rm kin} \, , \\
    \frac{\Omega_{\rm r}^0}{12\pi^2 \Omega_{\rm r}^{\rm end}}\left(\frac{H}{\m}\right)^2 \, , & f_{\rm kin}<f<f_{\rm end} \, , \, 
\end{cases}
\end{equation}
where

\begin{equation} \label{wperignniebfihbqwe}
f_{\rm end}=\frac{1}{2\pi} \left(\frac{\Omega_{\rm r}^0H_0^2H^2}{\Omega_{\rm r}^{\rm end}}\right)^{1/4} \, ,
\qquad
f_{\rm kin}=\frac{2}{\pi^2}\left(\frac{\Omega_{\rm r}^{0}H_0^2 H^2}{\Omega_{\rm r}^{\rm end}}\right)^{1/4}\left(\frac{1+\sqrt{1+36\alpha H^2/\m^2}}{2}\right)^{-\frac{1}{2}} \, ,
\end{equation}

\begin{equation} \label{adufbausdf}
    f_{\rm reh}=\frac{\left[\Omega_{\rm r}^0 (\Omega_{\rm r}^{\rm end})^3 H_0^2 H^2\right]^{1/4}}{4}\left(\frac{1+\sqrt{1+36\alpha H^2/\m^2}}{2}\right)^{-\frac{1}{2}}\,.
\end{equation}
\end{widetext}
\footnotetext{Note that, since we have set $a=1$ at the end of inflation instead of today as is customary, the numerical values of our $k$ differ from those of the usual comoving wavenumber. Equation \eqref{eq:f_vs_k} takes this into account.}
Note that the frequencies of the modes that cause the truncated peak, corresponding to hyperkination and kination, are always between $f_{\rm reh}$ and $f_{\rm end}$, given by Eqs. \eqref{adufbausdf} and \eqref{wperignniebfihbqwe}, respectively. The specific values depend on the Hubble parameter at the end of inflation $H$, the density parameter of radiation at the end of inflation $\Omega_{\rm r}^{\rm end}$ and $\alpha$. In order to give some indicative values, let us assume GUT scale inflation $H\simeq 10^{13}\,\text{GeV}$ and electroweak-scale reheating $\rho(\eta_{\rm reh})\simeq (200\,\text{GeV})^4$, corresponding to $\Omega_{\rm r}^{\rm end}=10^{-10}$. Changing $\alpha$ obviously leaves $f_{\rm end}$ unchanged. In Table~\ref{tab:frequencies}, we show $f_{\rm reh}$ and $f_{\rm end}$ for a few different $\alpha$. Note that they are larger than $f_{\rm BBN}$, as they should be.

\begin{table}[h]
\centering
\begin{tabular}{p{1cm}p{2.5cm}p{2.5cm}}
\textbf{$\alpha$} & $f_{\rm reh}$ & $f_{\rm end}$\\
\hline
$10^{30}$& $3.9\times 10^{-5}\,\text{Hz}$ & $4.4\times 10^{10}\, \text{Hz}$ \\
\hline
$10^{35}$ &$2.2\times 10^{-6}\,\text{Hz}$ & $4.4\times 10^{10}\, \text{Hz}$\\
\hline
$10^{40}$ & $1.2\times 10^{-7}\,\text{Hz}$ &$4.4\times 10^{10}\, \text{Hz}$\\
\end{tabular}
\caption{Values of the frequencies corresponding to reheating $f_{\rm reh}$ and the end of inflation $f_{\rm end}$ for different values of $\alpha$, given that $H= 10^{13}\,\text{GeV}$ and $\Omega_{\rm r}^{\rm end}=10^{-10}$.}
\label{tab:frequencies}
\end{table}

\subsection{Parameter space and detectability}
\label{sec:experiments}

In the present section, we put our model to the test and analyse the detectability of the generated spectrum of primordial GWs in the presence of a period of hyperkination after inflation. Since our analytical expression for the spectrum approximates very well its numerical counterpart, as can be seen from Fig.~\ref{fig:spectrum}, we use it in order to compare with the sensitivity curves of various detectors, namely LISA~\cite{Bartolo:2016ami,Caprini:2019pxz,LISACosmologyWorkingGroup:2022jok}, ET~\cite{Punturo:2010zz,Hild:2010id}, LVK observing runs O3 and O5~\cite{Harry:2010zz,VIRGO:2014yos,LIGOScientific:2014pky,LIGOScientific:2019lzm,KAGRA:2020tym}, SKA~\cite{Janssen:2014dka}, DECIGO~\cite{Kawamura:2006up,Kawamura:2011zz,Kawamura:2020pcg} and BBO~\cite{Harry:2006fi}. For each of them, we run a scan over the parameter space $\{\alpha, \Omega_{\rm r}^{\rm end},H\}$. The successful parameter space can be found in Fig.~\ref{fig:LIGOresults}. 

\begin{figure*}[h]
\centering
\includegraphics[width=0.6\textwidth]{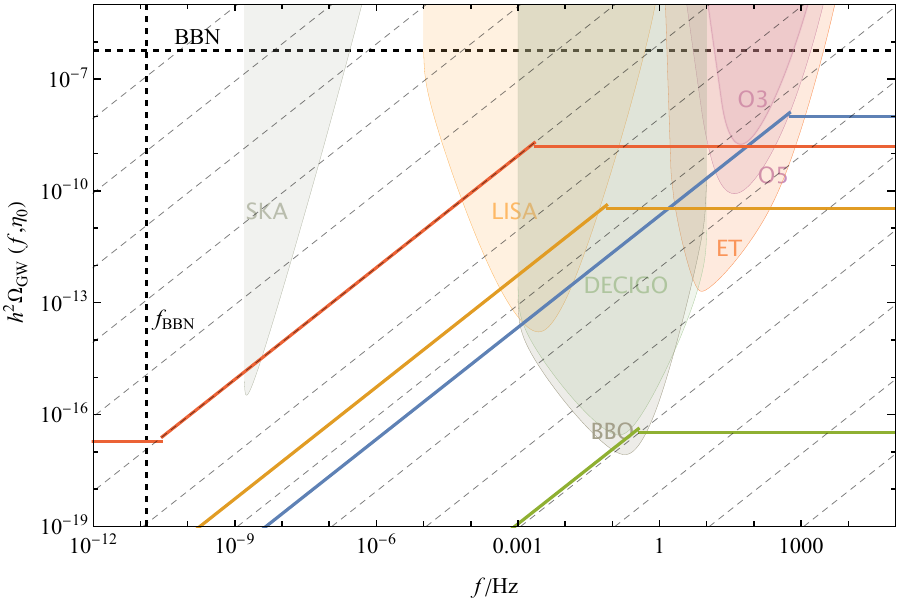}
\caption{A few different spectra superimposed with the PLIC curves of the GW experiments. The parameter values $\{N,H,\Omega_{\rm r}^{\rm end}\}$ are $\{17.5,\ 4.3\times  10^{11}\ \text{GeV},\ 10^{-12}\}$ for the blue curve, $\{25,\ 7.9\times 10^{11}\ \text{GeV},\ 10^{-9}\}$ for the orange curve, $\{20,\ 7.9\times 10^{10}\ \text{GeV},\ 10^{-5}\}$ for the green curve and $\{29.5,\ 1.7\times 10^{13}\ \text{GeV},10^{-8}\}$ for the red curve. We also show lines parallel (dashed gray) to the kination part of the spectrum. If not for the hyperkination period the spectra would violate the BBN bound.} 
\label{jdvasdbfiasdf}
\end{figure*}

Before we describe how the parameter space scan is performed, we comment on some bounds that need to be imposed. First, BBN should happen during the period of radiation domination. In other words, at (and below) the frequency associated with BBN, the spectrum needs to be in its lower plateau, \textit{i.e.}, $f_{\rm reh}>f_{\rm BBN}$, where $f_{\rm reh}$ is given by Eq.~\eqref{adufbausdf}. This imposes a bound on the maximum value $\alpha$ can take. Solving for $\alpha$ in Eq. \eqref{adufbausdf} gives
\begin{eqnarray} \label{adosjnanisdnfwer}
    \alpha&<&\frac{\m^2}{36 H^2}\left[\left(\frac{\sqrt{\Omega_{\rm r}^0(\Omega_{\rm r}^{\rm end})^3H_0^2H^2}}{8 f_{\rm BBN}^2}-1\right)^2-1\right]\nonumber \\
    &\simeq&\frac{\m^2 \Omega_{\rm r}^0(\Omega_{\rm r}^{\rm end})^3H_0^2}{2304 f_{\rm BBN}^4}=6.9\times 10^{85}(\Omega_{\rm r}^{\rm end})^3.
\end{eqnarray}

Importantly, we note here that the specific value we use for $f_{\rm BBN}$ in Eq. \eqref{adoijfbaibdfwef} comes from $T_{\rm BBN}= 1\, \text{MeV}$. However, recent studies \cite{Co:2021lkc,Oikonomou:2023bli} have shown that the stiff era is restricted to occur at temperatures $T>2.5\, \text{MeV}$. This means that the value in Eq. \eqref{adoijfbaibdfwef} would become a factor of $2.5$ larger, and the bound in Eq. \eqref{adosjnanisdnfwer} a factor of $0.026$ smaller. However, given that the available parameter space for $\alpha$ spans many order of magnitude (see Fig. \ref{fig:LIGOresults}), this change does not affect our results appreciably. Nevertheless, the reader should keep in mind that our bound $T> 1\, \text{MeV}$ is an approximate one.

\begin{figure}[h]
    \centering
        \includegraphics[width=0.4\textwidth]{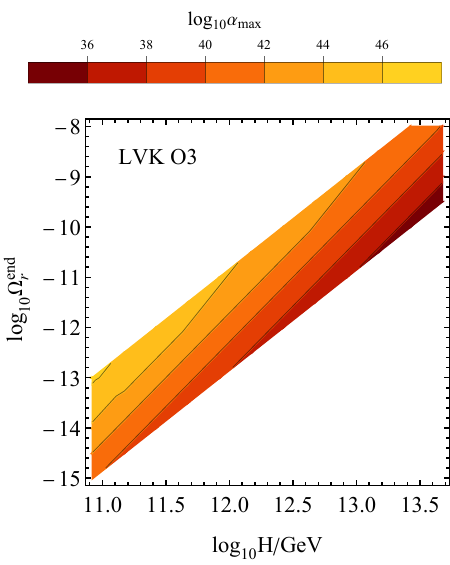}
        \caption{Parameter space of the theory excluded by LVK O3. For each value of $H$ and $\Omega_{\rm r}^{\rm end}$, there is a maximum value for $\alpha$, labelled $\alpha_{\rm max}$, above which the signal is observationally excluded.} 
\label{aisdjfijasndf}
\end{figure}

\begin{figure*}[h]
     \centering
    \begin{subfigure}[b]{0.3\textwidth}
         \centering
         \includegraphics[width=\textwidth]{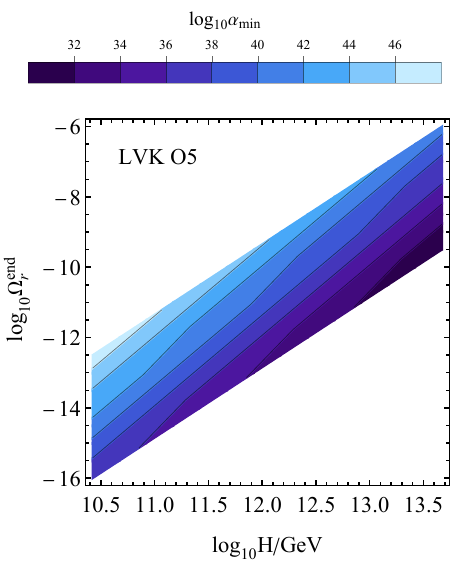}
     \end{subfigure}
     \begin{subfigure}[b]{0.3\textwidth}
         \centering
         \includegraphics[width=\textwidth]{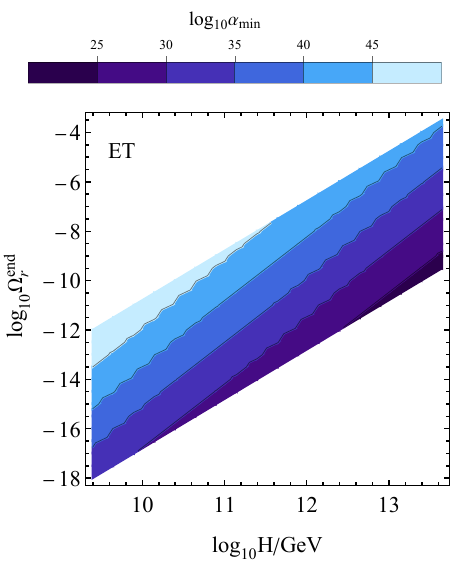}
     \end{subfigure}
     \begin{subfigure}[b]{0.3\textwidth}
         \centering
         \includegraphics[width=\textwidth]{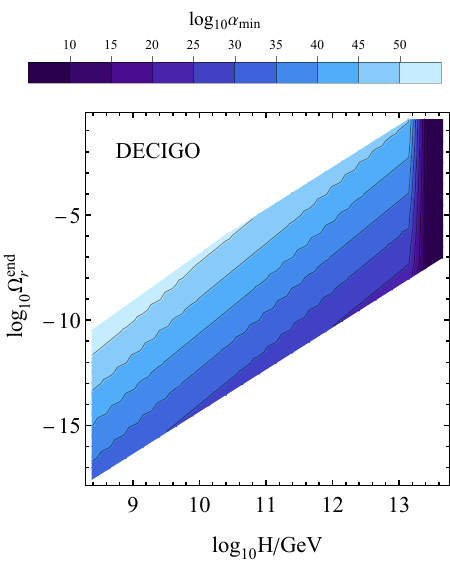}
     \end{subfigure}
     
     \bigskip
     \begin{subfigure}[b]{0.3\textwidth}
         \centering
         \includegraphics[width=\textwidth]{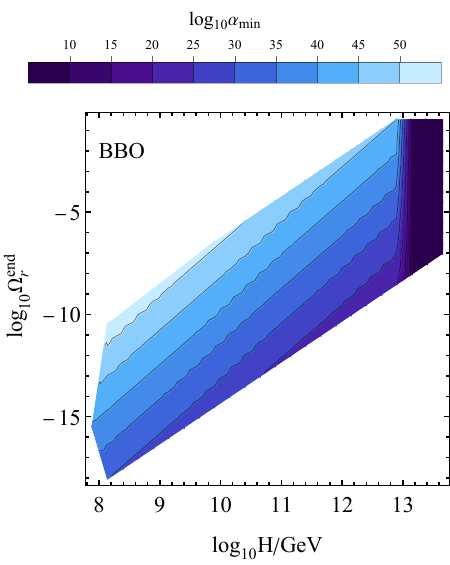}
     \end{subfigure}
     \begin{subfigure}[b]{0.3\textwidth}
         \centering
         \includegraphics[width=\textwidth]{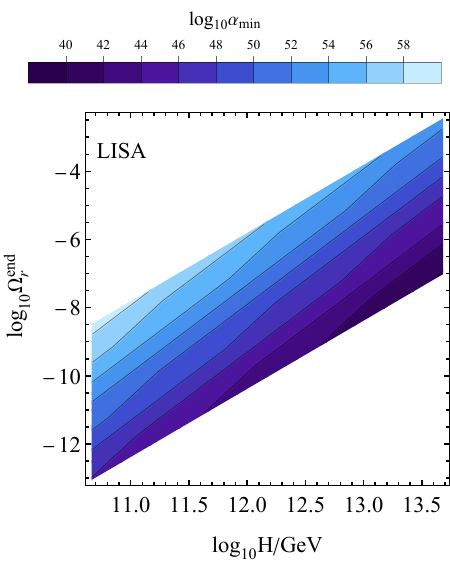}
     \end{subfigure}
     \begin{subfigure}[b]{0.3\textwidth}
        \centering
         \includegraphics[width=\textwidth]{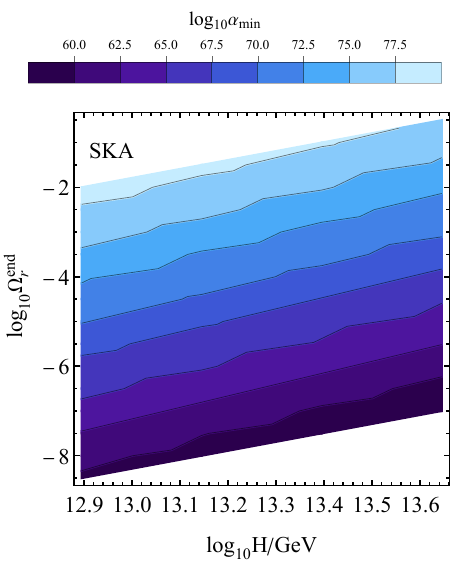}
     \end{subfigure}
     \caption{Parameter space of the theory for the minimum $\alpha$ such that the signal is detectable by LVK O5 (top left), ET (top middle), DECIGO (top right, BBO (bottom left), and LISA (bottom middle) and SKA (bottom right). For each value of $H$ and $\Omega_{\rm r}^{\rm end}$, there is a minimum value for $\alpha$, labelled $\alpha_{\rm min}$, above which the signal is always detectable (minus the excluded region in Fig. \ref{aisdjfijasndf} for LVK O5 and ET).}
      \label{fig:LIGOresults}
\end{figure*}

In addition, the GW energy density at BBN must be low enough not to disturb the standard results. Eqs.~\eqref{eq:Omega_GW} and \eqref{eq:deriving_rho_a4_eq} give $\Omega_{\rm GW}^{\rm BBN} = \Omega_{\rm GW}^0/\Omega_{\rm r}^0$, allowing us to translate the bound to into the GW energy density today, yielding \cite{Cyburt:2004yc}
\begin{equation} \label{eq:BBN_Omega_GW_bound}
    h^2 \Omega_{\rm GW}^0=\int \frac{\text{d}f}{f}h^2\Omega_{\rm GW}(f)<1.12\times 10^{-6} \, ,
\end{equation}
where $h\approx 0.7$ is the dimensionless Hubble constant. In practice, however, for all detectors except LVK O5 and ET, this bound is irrelevant. Indeed, it is sufficient to impose that the hyperkination plateau be below the minimum of LVK O3, the region excluded by now by LVK, which is below the BBN bound. Note that for LVK O5 and ET there exists some parameter space where the hyperkination plateau is between both limits. We take this into account in the scans by showing the excluded region from LVK O3 in Fig.~\ref{aisdjfijasndf}. There, for each value of $H$ and $\Omega_{\rm r}^{\rm end}$, we show the maximum value of $\alpha$, labelled $\alpha_{\rm max}$, below which the signal is not observationally excluded.

We can also impose an upper bound on the energy scale at the end of inflation. Using the slow-roll expression for the amplitude of the scalar power spectrum, we can write the Hubble parameter at CMB scales as
\begin{equation} \label{adiufbasdf}
    H_{\rm CMB}=\sqrt{\frac{\rho_{\rm CMB}}{3\m^2}}=\m \sqrt{A_s\frac{\pi^2 r}{2}} \, ,
\end{equation}
where $A_{\rm s}=2.1\times 10^{-9}$ \cite{Aghanim:2018eyx} and $r$ is the tensor-to-scalar ratio. The latest constraint on $r$ is $r<0.036$ \cite{BICEP:2021xfz}. The energy scale at the end of inflation is always lower than at CMB scales, so Eq.~\eqref{adiufbasdf} provides an upper bound on $H$ at the end of inflation,
\begin{equation} \label{eq:H_max}
    H< 4.7 \times 10^{13}\,\text{GeV} \, .
\end{equation}

Further, the plateau corresponding to radiation domination should be below the one corresponding to hyperkination, but this is not strictly guaranteed by our approximative spectrum if the kination period is short. To ensure this condition is satisfied, we impose
\begin{equation}
    \Omega_{\rm r}^{\rm end}<\frac{8}{\pi^2}\simeq 0.81 \, ,
\end{equation}
see Eq.~\eqref{aipsdbfbiuadsf}.

The logic for the parameter scan is as follows. We consider a grid in the $(H,\Omega_{\rm r}^{\rm end})$ plane, with the values of $H$ lying in the interval $[10^6,4.7\times 10^{13}]\,\text{GeV}$ and those of $\Omega_{\rm r}^{\rm end}$ lying in the interval $[10^{-20},0.81]$, both in steps of $0.5$ in logarithmic units. Then, for each point in the grid, we find the minimum value $\alpha_{\rm min}$, such that our spectrum is detectable by the specific experiment we are considering. Since the effect of increasing $\alpha$ (or, analogously, $N_{\rm hyp}$) is to stretch the flat region corresponding to hyperkination, if a signal is detectable for $\alpha_{\rm min}$, it will also be detectable for every $\alpha>\alpha_{\rm min}$. Note that for LVK O5 and ET, for a certain region in the $(H,\Omega_{\rm r}^{\rm end})$ plane, there is also a maximum value that $\alpha$ can take, see Fig.~\ref{aisdjfijasndf}. This limitation exists only for values where the height of the hyperkination plateau is above the minimum of the LVK O3 sensitivity curve.

In order to determine whether a signal can be detected, we compute the power-law integrated curves (PLIC) \cite{Thrane:2013oya} for each experiment. Then, for each set of parameters, we find the minimum $\alpha_{\rm min}$ such that the energy density spectrum is at least as large as the PLIC under consideration. An easy way to picture this procedure is to realise that the spectra with $\alpha=\alpha_{\rm min}$ are tangent to the PLICs. Increasing $\alpha$ increases the length of the hyperkination plateau, so if the spectrum is tangent to a PLIC, it will be above it for some frequency range if $\alpha>\alpha_{\rm min}$.

In Fig. \ref{jdvasdbfiasdf}, we show some example spectra with a large enough SNR, superimposed with the sensitivity curves for all considered experiments. In the same figure, we also show a grid of lines with the same slope as $\Omega_{\rm GW}^{\rm kin}(f,\eta_0)$ to showcase how in a setup with inflation being followed by usual kination most of the signals would violate the BBN bound. Hyperkination fixes this by truncating the spectrum and introducing a new plateau at high frequencies.

We report the results of parameter space scans as contour plots in Fig.~\ref{fig:LIGOresults}. There, for each pair $(H,\Omega_{\rm r}^{\rm end})$, we give the minimum $\alpha_{\rm min}$ such that the signal is detectable, for each experiment. We emphasize that the totality of the successful parameter space is contained in these figures. Besides the maximum value of $H$ from Eq.~\eqref{eq:H_max}, the parameter space is bounded at small $H$ by the BBN timing condition $f_{\rm reh}>f_{\rm BBN}$, at small $\Omega_{\rm r}^{\rm end}$ by the BBN energy density condition in Eq.~\eqref{eq:BBN_Omega_GW_bound} and the LVK O3 exclusion bound, and at large $\Omega_{\rm r}^{\rm end}$ by the requirement that the higher hyperkination plateau must reach the lower end of the sensitivity band for the given experiment.

We conclude that there is ample parameter space to accommodate detectability by all experiments. Indeed, as can be seen from Fig.~\ref{fig:LIGOresults}, for a Hubble parameter $H\lesssim 10^{13}$ GeV, somewhat below the GUT scale, and a reheating efficiency in the range of $10^{-15}\lesssim \Omega_{\rm r}^{\rm end}\lesssim 10^{-2}$, which can be easily accommodated by a variety of reheating mechanisms \cite{Felder:1998vq,Felder:1999pv,Dimopoulos:2018wfg,Opferkuch:2019zbd,Bettoni:2021zhq,Feng:2002nb,BuenoSanchez:2007jxm}, we can always find a detectable signal. We emphasize that the size of the parameter space is large, and there is no need for fine-tuning to obtain a detectable signal. Indeed, in Fig. \ref{fig:LIGOresults} we report the minimum value $\alpha$ has to take in order for the signal to be detectable. However, \textit{any} value of $\alpha$ larger than $\alpha_{\rm min}$ also leads to a detectable signal.

The value of $\alpha_{\rm min}$ is quite large for most experiments. This can be understood from Eq.~\eqref{eq:rho_end_vs_Nhyp}. Indeed, we can find a lower bound on $\alpha_{\rm min}$ by taking the limit $N_{\rm hyp}\ll 1$. It gives
\begin{equation}
    \alpha \rho_{\rm end}\simeq \frac{2\m^4}{3}N_{\rm hyp}.
\end{equation} 
Using a GUT energy scale $\rho_{\rm end}\sim 10^{-10}\m^4$, considering an almost non-existent period of hyperkination with $N_{\rm hyp}=0.1$, we obtain a rough lower bound $\alpha_{\rm min}\gtrsim 10^{10}$. As soon as we have a larger $N_{\rm hyp}$, $\alpha_{\rm min}$ grows exponentially with it. This is in line with our latest work \cite{Dimopoulos:2022rdp}, where we study quintessential inflation with an action of the form in Eq.~\eqref{eq:S_Jordan}. There, we find $\alpha\sim 10^{10}$ for successful quintessential inflation, without considerable hyperkination.

\section{Discussion and conclusions}
\label{sec:conclusions}

We have investigated  the spectrum of primordial gravitational waves (GWs) generated by cosmic inflation in a model where after inflation but before reheating we have a period when the Universe is dominated by the kinetic energy density of the inflaton scalar field $\phi$, when the field is characterised by both the usual quadratic kinetic term and also by a higher-order quartic kinetic term. This is natural in theories of quadratic $R+\alpha R^2$ gravity in the Palatini formalism, where in the Einstein frame the quartic kinetic term is proportional to $\alpha$, the coefficient of quadratic gravity. However, we can equally well envisage a $k$-inflation scenario where the kinetic term of the scalar field includes a term $\propto\alpha X^2$, where $X=\frac12\dot\phi^2$. 

This kinetically dominated period is divided into two parts. In the first part, the inflaton kinetic energy density is dominated by the higher-order kinetic term; a period which we call hyperkination. In the second part, the higher-order kinetic term becomes negligible and the inflaton kinetic energy density is dominated by the usual quadratic term; a period called kination. We have shown that, while kination is a stiff phase with barotropic parameter \mbox{$w=p/\rho=1$}, as is well known, hyperkination is not; the barotropic parameter during hyperkination is that of radiation $w=1/3$. As a result, the modes of inflation-generated primordial GWs which re-enter the horizon during hyperkination form a flat spectrum, in the same way as the modes which re-enter the horizon after reheating, in the usual radiation era. However, during usual kination, the GW spectrum is not flat but the GW density parameter per frequency logarithmic interval is \mbox{$\Omega_{\rm GW}(f)\propto f$}. This means that, for modes re-entering the horizon after inflation and before reheating, the GW signal is boosted. This boost corresponds to a truncated peak in the GW spectrum; truncated because the spectrum corresponding to hyperkination is flat but it can be of much larger amplitude than that corresponding to the eventual radiation era. Consequently, the period of kinetic domination (kination + hyperkination) can be made to last longer and the boosted spectrum to extend to lower frequencies without the danger of the production of excessive primordial GWs. In particular, the truncated spectrum can avoid the upper bound imposed by the requirement that Big Bang Nucleosynthesis (BBN) remains undisturbed. Thus, primordial GWs in all observable frequencies can be enhanced without a problem.

We have analytically and numerically studied thoroughly the inflationary production and the subsequent evolution of GW modes and obtained the resulting GW spectrum, linking it with the model parameters. The characteristic shape of the spectrum will be testable in the near future by forthcoming experiments, such as advanced LIGO-Virgo-KAGRA, LISA, DECIGO, BBO and ET, as depicted in Fig.~\ref{jdvasdbfiasdf}. If observed, such a spectrum can provide insight into the underlying theory, such as the energy scale of inflation, the reheating efficiency and the coefficient $\alpha$. The latter is directly related to the duration of the hyperkination phase. Indeed, when hyperkination lasts $N_{\rm hyp}$, then Eq.~\eqref{eq:rho_end_vs_Nhyp} suggests
\begin{equation}
\alpha=\frac{\m^4}{3\rho_{\rm end}}\,\exp(4N_{\rm hyp})\,,
\label{alphaNhyp}
\end{equation}
where \mbox{$\rho_{\rm end}=3H^2\m^2$} is the energy density at the end of inflation, and $H$ is the corresponding Hubble scale. Typically, inflation is at the scale of grand unification, which implies \mbox{$H^2\sim 10^{-10}\,\m^2$}. In this case, the above suggests that \mbox{$e^{N_{\rm hyp}}\sim 10^{-3}\alpha^{1/4}$}, which means that
\begin{equation}
    N_{\rm hyp}\simeq 10\Rightarrow\alpha\sim 10^{26}.
\end{equation}
Note that, in the usual Starobinsky $R^2$ inflation we have \mbox{$\alpha=1.1\times 10^9$}. Such large values of alpha are non-perturbative, but this is no more a problem in our setup than it is in Starobinsky gravity.

Important information can also be deduced by the amplitude of the truncated peak corresponding to hyperkination. Indeed, Eq.~\eqref{eq:Omega_GW_spectrum_today} suggests that the value of the GW spectrum on the hyperkination plateau is given by
\begin{equation}
\Omega_{\rm GW}^{\rm hyp}=\frac{1}{12\pi^2}\frac{\Omega_r^0}{\Omega_r^{\rm end}}
\left(\frac{H}{\m}\right)^2\,,
\label{OmegaGWhyp}
\end{equation}
where \mbox{$\Omega_r^0\simeq 10^{-4}$} is the density parameter of radiation at present and \mbox{$\Omega_r^{\rm end}$} is the density parameter of radiation at the end of inflation, also called reheating efficiency, because the larger it is the sooner reheating takes place. As discussed, in order not to destabilise BBN, we need \mbox{$\Omega_{\rm GW}^{\rm hyp}<10^{-6}$}. Thus, we obtain a lower bound on the reheating efficiency as \mbox{$\Omega_r^{\rm end}>(H/\m)^2$}. Typically for inflation we have \mbox{$H^2\sim 10^{-10}\m^2$}, which implies \mbox{$\Omega_r^{\rm end}>10^{-10}$}.

In an effort to stay generic, we have not considered a specific mechanism for producing the radiation which eventually reheats the Universe. We note however, that a number of such mechanisms exist, such as instant preheating~\cite{Felder:1998vq,Felder:1999pv}, curvaton reheating~\cite{Feng:2002nb,BuenoSanchez:2007jxm} or Ricci reheating~\cite{Dimopoulos:2018wfg,Opferkuch:2019zbd,Bettoni:2021zhq} to name but some. It is even possible to avoid introducing additional degrees of freedom and consider that reheating occurs due to the dissipating properties of the inflaton field itself, as discussed in Ref.~\cite{Dimopoulos:2019gpz}, where such processes become negligible after inflation.

Additional important information can be obtained by the observation of the frequency of the knee in the GW spectrum, shown in Figs.~\ref{fig:spectrum} and \ref{jdvasdbfiasdf}, which is given by $f_{\rm kin}$ in Eq.~\eqref{wperignniebfihbqwe}. Combining this with Eq.~\eqref{OmegaGWhyp}, in the large $N_{\rm hyp}$ limit, 
we obtain
\begin{equation}
\frac{f_{\rm kin}}{(\Omega_{\rm GW}^{\rm hyp})^{1/4}}=\frac{2}{\pi^{3/2}}\sqrt{\frac{2}{3}}\rho_0^{1/4}\,
\alpha^{-1/4}\sqrt{\frac{\m}{H}}\,,
\label{fkneeOmegaGW}
\end{equation}
Where \mbox{$\rho_0=3H_0^2\m^2$} is the energy density of the Universe at present. Putting the numbers in the above, we find
\begin{equation}
\left(\frac{f_{\rm kin}}{\rm Hz}\right)
\left(\frac{\Omega_{\rm GW}^{\rm hyp}}{10^{-6}}\right)^{-1/4}\sim
10^{12}\,\alpha^{-1/4}\sqrt{\frac{\m}{10^3\,H}}\,.
\label{fkneeOmegaGWfin}
\end{equation}
Observations might provide the values of the left-hand-side of the above, which means that $\alpha$ could be estimated provided $H$ is known (e.g. \mbox{$H^2\sim 10^{-10}\m^2$} for inflation at the grand unified energy scale).

In Fig.~\ref{fig:LIGOresults} we display our findings with respect to observability by different missions, such as LVK 05, ET, BBO, LISA DECIGO and SKA. There, we show the minimum value $\alpha$ has to take in order for the spectrum to be detectable. Above this value, which we label $\alpha_{\rm min}$, the spectrum is always detectable. We see that observability requires that the reheating efficiency is smaller the lower the inflation energy scale is (the lower $H$ is). Also, the values of $\alpha_{\rm min}$ are larger for large inflationary energy scales. For LVK 05 and LISA we find that observability requires \mbox{$\alpha_{\rm min}\sim 10^{30-60}$}, while for ET, BBO and DECIGO the numbers are smaller \mbox{$\alpha_{\rm min}\sim 10^{10-50}$}. For the reheating efficiency, we find that observability requires that the density parameter of radiation at the end of inflation is \mbox{$\Omega_{\rm GW}^{\rm end}\gtrsim 10^{-16}$}, a value which may increase up to unity or so in the case of ET, BBO or DECIGO. Such a high reheating efficiency implies that the kinetic regime is very small or even non-existent (prompt reheating). This is possible because, the ET, BBO and DECIGO might be able to detect very faint signals at frequencies higher than LISA, which means that they could even marginally observe the flat GW spectrum generated by the usual radiation era (no kinetic epoch). This is why there is a region (for ET, BBO and DECIGO) when $H$ is large (\mbox{$H\sim 10^{13}\,$GeV}) where suddenly $\alpha$ can be very small (or even zero). The parameter space for this is very small though.

We conclude that, with our mechanism, the observability of primordial GWs is much enhanced compared to traditional models. We obtained concrete predictions involving $H$, $\alpha$ and the reheating efficiency in the case the characteristic form of the GW spectrum---a truncated peak---is indeed observed. Observation of the primordial GW signal would not only confirm another prediction of cosmic inflation but would also be a tantalising hint towards the quantum nature of gravity, which is behind the assumption of the Bunch-Davies vacuum in Eq.~\eqref{eq:Bunch-Davies}. Forthcoming GW observations may reveal new and surprising details about the physics of inflation and fundamental physics in general. Our work serves to explore such a possibility.

\section*{Acknowledgments}

We thank Maciek Kierkla, Ville Vaskonen and Hardi Veerm\"ae for useful comments and discussions. SSL was supported by the FST of Lancaster University. KD was supported, in part, by the Lancaster-Manchester-Sheffield Consortium for Fundamental Physics under STFC grant: ST/T001038/1. AK and ET were supported by the Estonian Research Council grants PSG761 and PRG1055 and by the EU through the European Regional Development Fund CoE program TK133 ``The Dark Side of the Universe".  For the purpose of open access, the authors have applied a Creative Commons Attribution (CC BY) licence to any Author Accepted Manuscript version arising.

\appendix

\section{\boldmath A toy model for a drastic change of $\alpha$ at the end of inflation}
\label{appx:toy}

The coefficient $\alpha$ parametrising quadratic gravity can experiece a drastic change at the end of inflation if it is a function of a degree of freedom which changes rapidly at that time. For example, if inflation takes place at the energy of grand unification, as is typically the case, then this degree of freedom could be the Higgs field $\chi$ of a Grand Unified Theory (GUT). If the breaking of grand unification takes place via spontaneous symmetry breaking, then the expectation value of $\chi$ changes from zero to $M\sim 10^{16}\,$GeV. 

A toy model example of the inflaton potential, which leads to the GUT phase transition but still retains the runaway nature assumed in this work is
\begin{eqnarray}
    &&V(\varphi,\chi)=\frac14\lambda(\chi^2-M^2)^2\nonumber \\
    &&+\left\{
    \begin{array}{ll}
\frac12(m^2+g^2\chi^2)(\varphi^2+\mu^2) & ,\;\varphi<0\\
\frac12(m^2+g^2\chi^2)
\,\mbox{\Large $\frac{\mu^6}{\varphi^4+\mu^4}$} & ,\;\varphi\geq 0\end{array}\right.\,,
\label{toyV}
\end{eqnarray}
where $m$ and $\mu$
are mass scales with \mbox{$0<\mu\ll m<M$}
and \mbox{$\lambda, g\lesssim{\cal O}(1)$}.
By taking $\lambda=0=g$,
we recover the \mbox{$(n,q)=(2,4)$} case of the quintessential inflation potential in an $R+R^2$ Palatini modified gravity theory, which was investigated in Ref.~\cite{Dimopoulos:2020pas}. This potential, in turn, is a minor modification of the original quintessential inflation potential in Ref.~\cite{Peebles:1998qn}. Switching $\lambda$ and $g$ on, and considering the limit \mbox{$|\varphi|\gg\mu$} with \mbox{$\varphi<0$}, we obtain the original hybrid inflation potential \cite{Linde:1993cn}. 

Let us first consider standard gravity without an $R^2$ term. In the beginning, \mbox{$\varphi\ll-\mu$}. Then the effective mass-squared of the GUT Higgs field $\chi$ is positive, which sends $\chi$ to zero. The scalar potential then becomes
\begin{equation}
V=\frac12m^2\varphi^2+\frac14\lambda M^4\,.
\label{Vinf}
\end{equation} 
When the constant term dominates, we have an inflationary plateau. The effective mass-squared of the GUT Higgs field is \mbox{$m_{{\rm eff}\chi}^2=g^2\varphi^2-\lambda M^2$}. Thus, $m_{{\rm eff}\chi}^2$ is positive as long as \mbox{$|\varphi|>|\varphi_c|$}, where 
\mbox{$\varphi_c\equiv -(\sqrt\lambda/g)M$}, where for simplicity we assume \mbox{$|\varphi_c|\gg\mu$}. Inflation ends when \mbox{$\varphi=\varphi_c$}, which triggers a phase transition that sends the GUT Higgs field towards its vacuum expectation value (VEV) \mbox{$\chi=M$}, in which case \mbox{$m_{{\rm eff}\chi}^2=2\lambda M^2$}. At this time, the effective mass-squared of the inflaton field becomes \mbox{$m_{{\rm eff}\varphi}^2=g^2M^2>0$}, when the inflaton is still negative \mbox{$\varphi_c<\varphi<0$}. This propels the inflaton to the origin.

When $\varphi$ becomes positive, it free-falls in its steep runaway potential. In the limit \mbox{$\varphi\gg\mu$}, the potential is
\begin{equation}
V=\frac{\frac12 g^2M^2\mu^6}{\varphi^4}\,,
\label{VQ}
\end{equation}
where we assumed 
\mbox{$gM>m$}.
The above inverse quartic potential can indeed work not as tracker quintessece, as in the original quintessential inflation model \cite{Peebles:1998qn}, but as a freezing-thawing quintessence, which unfreezes at present provided the mass-scale $(\frac12 g^2M^2\mu^6)^{1/8}$ is of the correct magnitude to satisfy the coincidence requirement. Inflation, however, as described above would not work. Indeed, the original hybrid inflation model
of Ref.~\cite{Linde:1993cn}, which is characterised by the inflationary potential in Eq.~\eqref{Vinf}, produces a blue spectral index for the scalar curvature perturbation.

As shown in Ref.~\cite{Dimopoulos:2020pas}, things change when we embed the above model in $R+R^2$ Palatini modified gravity. We assume that $\lambda$ is small enough, such that the potential in Eq.~\eqref{Vinf} during inflation is $V\simeq\frac12 m^2\varphi^2$. Then, the inflationary plateau is due to the quadratic gravity term, which flattens the potential and creates the inflationary plateau with
$U_{\rm inf}\simeq \m^4/4\alpha$ as discussed in Sec.~\ref{sec:quartic_kinetic_terms}. As mentioned, the scenario with $\lambda=0=g$ was investigated in Ref.~\cite{Dimopoulos:2020pas}, which found that successful quintessential inflation in achieved if $m\sim 10^{13}\,$GeV and 
$(\frac12 g^2M^2\mu^6)^{1/8}\sim 10\,$GeV, which means 
$\mu\sim g^{-1/3} 10^{-4}\,$GeV. The assumption $g>m/M\sim 10^{-3}$ suggests that $\mu\lesssim{\cal O}({\rm MeV})$. 

In Ref.~\cite{Dimopoulos:2020pas} it was shown that for successful quintessential inflation with this model we need
\mbox{$\alpha\sim 10^8$}, so that \mbox{$U_{\rm inf}^{1/4}\sim 10^{16}\,$GeV}. The canonical inflaton field rolls down the Palatini inflationary plateau $U_{\rm inf}$ until it triggers the GUT phase transition and sends the GUT Higgs field to its VEV. Then, the potential $V$ is reduced drastically so that the system exits the Palatini plateau and \mbox{$U\simeq V$}. 

The change of the expectation value of the GUT Higgs field $\chi$ at the phase transition not only terminates inflation but may also affect the value of $\alpha$ provided the latter depends on $\chi$.
Indeed, suppose that
\begin{equation}
    \alpha=\alpha(\chi)=\alpha_0\,e^{\kappa\chi/M}\,,
    \label{alphachi}
\end{equation}
where \mbox{$\kappa={\cal O}(10)$} is a coefficient and \mbox{$\alpha_0\sim 10^8$}. Before the phase transition, \mbox{$\chi=0$} and \mbox{$\alpha=\alpha_0\sim 10^8$}. After the phase transition, \mbox{$\chi=M\sim 10^{16}\,$GeV} and \mbox{$\kappa\chi/M\lesssim 10^2$}. As a consequence, $\alpha$ becomes huge. Indeed, for the range
\mbox{$\kappa=5-166$} we find
\mbox{$\alpha\sim 10^{\text{10--80}}$}, which comfortably includes the values considered in Fig.~\ref{fig:LIGOresults}. Note that $\alpha$ should not depend on the inflaton field, \mbox{$\alpha\neq\alpha(\varphi)$}, because the latter changes substantially during kination and hyperkination, while $\alpha$ is taken to be constant.

Finally, it must be pointed out that the period of hyperkination in the post-inflationary history would modify the treatment of Ref.~\cite{Dimopoulos:2020pas} somewhat. As a result, the value of $\mu$ for successful coincidence might change, but this is beyond the scope of the present work.

\section{Numerical solutions}
\label{sec:numerics}

To check our analytical results, we solve numerically the time evolution of the background composed of the field and radiation and the GW mode functions. The full set of equations reads
\begin{eqnarray} 
    &&\qty(1+3\alpha\frac{\dot{\phi}^2}{\m^4})\ddot{\phi} + 3\qty(1+\alpha\frac{\dot{\phi}^2}{\m^4}) H \dot{\phi} = 0 \, , \nonumber \\
    &&\dot{\rho}_f = -3H\rho_f(1+w_f) \,,\quad 3H^2\m^2 = \rho_\phi + \rho_f\, , \label{eq:collected_eoms} \\
    &&\rho_\phi = \frac{1}{2}\qty[1+\frac{3}{2}\alpha\frac{\dot{\phi}^2}{\m^4}]\dot{\phi}^2 \, ,\,\, h^s_k{}'' + 2\frac{a'}{a}h^s_k{}' + k^2 h_k^s = 0. \nonumber
\end{eqnarray}
Many of the variables vary by orders of magnitude during cosmic evolution. To make numerics easier, we define new, rescaled variables $x$, $y$, and $Z$, a new time variable $s$, and a constant $s_0$ through
\begin{eqnarray}
    \label{eq:scaled_variables}
    &&\dot{\phi} = \m^2 \alpha^{-1/2}e^{-s_0-s+x} \, , \quad
    \rho_f = \m^4 \alpha^{-1}e^{-2s_0-2s+y} \, , \nonumber \\
    &&H = \m \alpha^{-1/2} Z e^{-s_0-s} \, , \quad
    s_0 = -\ln\qty(2\sqrt{\alpha}H_0/\m) \, , \nonumber \\
    &&\dd t = \m^{-1}\sqrt{\alpha}e^{s_0+s}\dd s \, ,
\end{eqnarray}
where $H_0$ is the initial Hubble parameter. Definitions in Eq.~\eqref{eq:scaled_variables} are chosen to ensure the new numerical quantities remain of order one throughout the computation. The equations of motion become
\begin{eqnarray}
    \label{eq:rescaled_eoms}
    \accentset{\circ}{x} &=& 1-\frac{3Z\qty(1+e^{-2s_0-2s+2x})}{1+3e^{-2s_0-2s+2x}} \, ,\,\, \dotS{y} = 2-3Z\qty(1+w_f)\,,\nonumber \\
    &&3Z^2 = \frac{1}{2}\qty(1+\frac{3}{2}e^{-2s_0-2s+2x})e^{2x} + e^y \, , \nonumber\\
    &&\ddotS{h}_k + (3Z-1)\dotS{h}_k + \frac{k^2}{\m^2 a^2}\alpha e^{2s+2s_0}h_k = 0 \, ,
\end{eqnarray}
where a circle over a variable indicates a derivative with respect to the new time variable~$s$.

The initial conditions for the field velocity and fluid energy density are set as described in the text, engineered to match a desired end-of-inflation Hubble parameter $H_{\rm end}$, duration of hyperkination $N_{\rm hyp}$, and initial radiation energy density fraction $\Omega_{\rm r}^{\rm end}$. We then follow their evolution from the end of inflation until the BBN temperature is reached, see Fig.~\ref{fig:w_and_rho}. The gravitational wave modes are evolved from their frozen super-Hubble state in Eq.~\eqref{adufubauisdf} starting somewhat before they re-enter the Hubble radius, until somewhat after the re-entry, after which they are taken to behave as radiation. To get the mode energy density, we use the first equation in Eq.~\eqref{eq:rho_GW}---as explained in the text, the error related to regularization is negligible for all relevant modes. Iterated over a number of modes, this produces the spectra in Fig.~\ref{fig:spectrum}.

\section{Mode function matching} \label{appendixA}
In this appendix, we report the more technical results concerning the mode function matching at the transition between the different cosmological eras. We start with the transition from inflation to hyperkination, which takes place at $\eta_{\rm end}$. During the hyperkination, the Mukhanov Sasaki equation reads (see Eqs.~\eqref{eq:v_eom} and \eqref{eq:a_prime_prime})
\begin{equation}
    v_k^s{}''+k^2 v_k^s=0 \, .
\end{equation}
The solution is simply a superposition of plane waves,
\begin{equation} \label{adifbahsudf}
    v_k^s(\eta)=\frac{1}{\sqrt{2k}}\left(\alpha_{+}e^{-ik\eta}+\alpha_{-}e^{ik\eta}\right) \, .
\end{equation}
Matching this to the standard slow-roll result (see the first line of Eq.~\eqref{eq:v_k_cases}) at $\eta_{\rm end}$ gives
\begin{eqnarray}
    &&e^{i\frac{\pi}{4}\left(1+2\nu\right)}\sqrt{\frac{\pi}{2}}\sqrt{x_{\rm end}}H_{\nu}^{(1)}(x_{\rm end})\nonumber \\
    &&=\alpha_{+}e^{ik\abs{\eta_{\rm end}}}+\alpha_{-}e^{-ik\abs{\eta_{\rm end}}} \, ,
\end{eqnarray}
where $x_{\rm end}\equiv k\abs{\eta_{\rm end}}$ and we dropped the subindex $I$ from $\nu$. Matching the derivatives gives
\begin{eqnarray}
    &&i\sqrt{\frac{\pi}{2}}e^{i\frac{\pi}{4}\left(1+2\nu\right)}\Big[\frac{1}{\sqrt{x_{\rm end}}}\Big(\frac{1}{2}+\nu\Big)H_{\nu}^{(1)}(x_{\rm end})\nonumber\\
    &&-\sqrt{x_{\rm end}}H_{\nu+1}^{(1)}(x_{\rm end})\Big]\nonumber \\
    &&= -\alpha_{+}e^{ik\abs{\eta_{\rm end}}}+\alpha_{-}e^{-ik\abs{\eta_{\rm end}}} \, .
\end{eqnarray}
Summing (subtracting) both expressions, we obtain
\begin{eqnarray}
    &&\alpha_{\mp}=\frac{e^{i\frac{\pi}{4}\Big(1+2\nu\Big)\pm ix_{\rm end}}}{2}\sqrt{\frac{\pi}{2}}\Big[H_{\nu}^{(1)}(x_{\text{end}})\Big(\sqrt{x_{\text{end}}}\nonumber \\
    &&\pm\frac{i}{\sqrt{x_{\text{end}}}}(\nu+\frac{1}{2})\Big)\mp i\sqrt{x_{\text{end}}}H_{\nu+1}^{(1)}(x_{\text{end}})\Big] \, .
\end{eqnarray}

We now take the super-Hubble (small argument) limit $x_{\rm end}\ll 1$. Noting that the leading contributions come from the terms proportional to $H_{\nu}^{(1)}(x_{\text{end}})/\sqrt{x_{\text{end}}}$ and $ H_{\nu+1}^{(1)}(x_{\text{end}})\sqrt{x_{\text{end}}}$, it reads
\begin{equation} 
    \alpha_{\mp}=\pm \frac{2^{\nu-1}e^{i\frac{\pi}{4}\left(1+2\nu\right)}}{\sqrt{2\pi}}\left(\frac{1}{2}-\nu\right)\Gamma(\nu)\frac{1}{(k\abs{\eta_{\text{end}}})^{\nu+\frac{1}{2}}} \, .
\end{equation}
Using $\nu=3/2+\epsilon$, this expression can be further simplified to
\begin{equation} \label{asdifubasidf}
    \alpha_{\mp}=\pm \frac{2^{\epsilon-1}e^{i\pi\epsilon/2}\Gamma(3/2+\epsilon)}{\Gamma(3/2)}\left(\frac{H}{k}\right)^{2+\epsilon} \, .
\end{equation}
For pure de Sitter, with $\epsilon \to 0$, we obtain
\begin{equation}
    \alpha_{\mp }=\pm \frac{H^2}{2k^2} \, .
\end{equation}

We continue with the transition from hyperkination to kination, which takes place at $\eta_{\rm kin}$. During kination, the Mukhanov--Sasaki equation takes the form
\begin{equation}
    v_k^s{}''+\left[k^2 -\frac{1}{4\left[\eta-\frac{\eta_{\rm kin}}{2}+\frac{1}{H}\right]^2}\right]v_k^s=0 \, .
\end{equation}
Making the change of variables $y\equiv k\left(\eta-\eta_{\rm kin}/2+1/H\right)$ (where $y=kz$ in the notation of Eq.~\eqref{eq:z}) and redefining the mode functions as $g=\sqrt{y}v$, this equation can be recast as a Bessel equation with $\nu=0$ (see Eq.~\eqref{asiphdbfajsdf}). Thus, the solution reads
\begin{eqnarray} 
    v_k^s(\eta)&&=\sqrt{\frac{\pi}{4k}}\sqrt{y}\Big[e^{-i\pi/4}\beta_{+}(k)H_{0}^{(2)}(y)\nonumber \\
    &&+e^{i\pi/4}\beta_{-}(k)H_0^{(1)}(y)\Big] \, ,
\end{eqnarray}
where the overall constant and phase has been chosen such that the mode functions have a simple sub-Hubble ($y\gg 1$) limit, as discussed below Eq.~\eqref{eq:v_k_cases}. We match this equation (and its derivative) with Eq.~\eqref{adifbahsudf} (and its derivative) at time $\eta_{\rm kin}$, \textit{i.e.}, at
\begin{equation}
    y_{\text{kin}}\equiv y(\eta_{\rm kin})=\frac{k}{2}\left(\eta_{\rm kin}+\frac{2}{H}\right)\simeq \frac{k\eta_{\rm kin}}{2} \, ,
\end{equation}
where we have taken into account that $\eta_{\rm kin}\gg\eta_{\rm end}$. To avoid clutter we also define $r\equiv e^{i\pi/4}\sqrt{\pi/2}$. Equating the mode functions gives
\begin{eqnarray} 
    &&\alpha_{+}e^{-ik\eta_{\rm kin}}+\alpha_{-}e^{ik\eta_{\rm kin}}\nonumber \\
    &&=\sqrt{y_{\rm kin}}\left[r^{*}\beta_{+}H_0^{(2)}(y_{\rm kin})+r\beta_{-}H_0^{(1)}(y_{\rm kin})\right] \, ,\label{ihqerbguhbasdbf}
\end{eqnarray}
while doing so for the derivatives gives
\begin{eqnarray} 
    &&i\left(-\alpha_{+}e^{-ik\eta_{\rm kin}}+\alpha_{-}e^{ik\eta_{\rm kin}}\right)\nonumber \\
    &&=\frac{1}{2\sqrt{y_{\rm kin}}}\left[r^{*}\beta_{+}H_0^{(2)}(y_{\rm kin})+r\beta_{-}H_0^{(1)}(y_{\rm kin})\right] \nonumber \\
    &&+\sqrt{y_{\rm kin}}\left[r^{*}\beta_{+}\frac{\text{d}H_0^{(2)}}{\text{d}y}(y_{\rm kin})+r\beta_{-}\frac{\text{d}H_0^{(1)}}{\text{d}y}(y_{\rm kin})\right] \, .\label{aipdfbhsdfsdf}
\end{eqnarray}
Now, using Eq.~\eqref{ihqerbguhbasdbf} in Eq.~\eqref{aipdfbhsdfsdf} allows us to rewrite the latter as
\begin{equation} \label{aisdbfhausbdf}
\begin{aligned}
    &\left[\alpha_{+}\left(-i-\frac{1}{2y_{\rm kin}}\right)e^{-ik\eta_{\rm kin}}+\alpha_{-}\left(i-\frac{1}{2y_{\rm kin}}\right)e^{ik\eta_{\rm kin}}\right]\\
    &=\sqrt{y_{\rm kin}}\left[r^{*}\beta_{+}\frac{\text{d}H_0^{(2)}}{\text{d}y}(y_{\rm kin})+r\beta_{-}\frac{\text{d}H_0^{(1)}}{\text{d}y}(y_{\rm kin})\right] \, .
\end{aligned}
\end{equation}
In order to obtain $\beta_{-}$ ($\beta_{+}$), we multiply Eq.~\eqref{aisdbfhausbdf} by $H_{0}^{(2)}(y_{\rm kin})$ ($H_0^{(1)}(y_{\rm kin})$) and Eq.~\eqref{ihqerbguhbasdbf} by $\text{d}H_0^{(2)}/\text{d}y$ ($\text{d}H_0^{(1)}/\text{d}y$), subtract the latter from the former and use the Wronskian of the Hankel functions. The results read
\begin{equation}
\begin{aligned}
    &\beta_{-}=e^{-i\pi/4}\frac{\sqrt{\pi y_{\rm kin}}}{i2\sqrt{2}}\Biggl\{H_{0}^{(2)}(y_{\rm kin})\Big[\alpha_{+}\left(-i-\frac{1}{2y_{\rm kin}}\right)e^{-ik\eta_{\rm kin}}\\
    &+\alpha_{-}\left(i-\frac{1}{2y_{\rm kin}}\right)e^{ik\eta_{\rm kin}}\Big]\\
    &+H_{1}^{(2)}(y_{\rm kin})\left(\alpha_{+}e^{-ik\eta_{\rm kin}}+\alpha_{-}e^{ik\eta_{\rm kin}}\right)\Biggl\}
\end{aligned}
\end{equation}
and
\begin{equation}
\begin{aligned}
    &\beta_{+}=-e^{i\pi/4}\frac{\sqrt{\pi y_{\rm kin}}}{i2\sqrt{2}}\Biggl\{H_{0}^{(1)}(y_{\rm kin})\Big[\alpha_{+}\left(-i-\frac{1}{2y_{\rm kin}}\right)e^{-ik\eta_{\rm kin}}\\
    &+\alpha_{-}\left(i-\frac{1}{2y_{\rm kin}}\right)e^{ik\eta_{\rm kin}}\Big]+H_{1}^{(1)}(y_{\rm kin})\Big(\alpha_{+}e^{-ik\eta_{\rm kin}}\\
    &+\alpha_{-}e^{ik\eta_{\rm kin}}\Big)\Biggl\} \, .
\end{aligned}
\end{equation}
Noting that $\alpha_{+}=-\alpha_{-}$, these expressions can be rewritten as
\begin{equation}
\begin{aligned}
    &\beta_{-}=e^{-i\pi/4}\sqrt{\frac{\pi y_{\rm kin}}{2}}\alpha_{-}\Biggl\{H_{0}^{(2)}(y_{\rm kin})\Big[\cos{(k\eta_{\rm kin})}\\
    &-\frac{1}{2y_{\rm kin}}\sin{(k\eta_{\rm kin})}\Big]+H_{1}^{(2)}(y_{\rm kin})\sin{(k\eta_{\rm kin})}\Biggl\}
\end{aligned}
\end{equation}
and
\begin{equation}
\begin{aligned}
    &\beta_{+}=-e^{i\pi/4}\sqrt{\frac{\pi y_{\rm kin}}{2}}\alpha_{-}\Biggl\{H_{0}^{(1)}(y_{\rm kin})\Big[\cos{(k\eta_{\rm kin})}\\
    &-\frac{1}{2y_{\rm kin}}\sin{(k\eta_{\rm kin})}\Big]+H_{1}^{(1)}(y_{\rm kin})\sin{(k\eta_{\rm kin})}\Biggl\} \, .
\end{aligned}
\end{equation}

We can now take the super-Hubble limit $k\eta_{\rm kin}\ll 1$. Using $k\eta_{\rm kin}=2y_{\rm kin}$, the term in brackets multiplying $H_0^{(1,2)}(y_{\rm kin})$ cancels out, and we obtain the result
\begin{equation} \label{asdifubasdf}
    \beta_{\pm}=2i e^{\pm i\pi/4}\alpha_{-}\sqrt{\frac{k\eta_{\rm kin}}{\pi}} \, ,
\end{equation}
where $\alpha_{-}$ is given by Eq. \eqref{asdifubasidf}. Note that
\begin{equation} \label{asdifubabuhsdf}
    \beta_{+}=i\beta_{-} \, .
\end{equation}
For pure de Sitter, we have the simplified expression
\begin{equation}
    \beta_{\pm}=ie^{\pm i \pi/4}\left(\frac{H}{k}\right)^2\sqrt{\frac{k \eta_{\rm kin}}{\pi}} \, .
\end{equation}

Finally, we consider the transition from kination to the radiation-dominated era at $\eta_{\rm reh}$. During the latter, the Mukhanov--Sasaki equation is identical to the one corresponding to hyperkination,
\begin{equation}
    v_k^s{}''+k^2v_k^s=0 \, ,
\end{equation}
the solution to which reads
\begin{equation} 
    v_k^s(\eta)=\frac{1}{\sqrt{2k}}\left(\gamma_{+}e^{-ik\eta}+\gamma_{-}e^{ik\eta}\right) \, .
\end{equation}
The matching conditions at $\eta_{\rm reh}$ now read
\begin{eqnarray}
    &&\sqrt{y_{\rm reh}}\left[r^{*}\beta_{+}H_0^{(2)}(y_{\rm reh})+r\beta_{-}H_0^{(1)}(y_{\rm reh})\right]\nonumber \\
    &&=\left(\gamma_{+}e^{-ik\eta_{\rm reh}}+\gamma_{-}e^{ik\eta_{\rm reh}}\right)
\end{eqnarray}
and
\begin{equation}
\begin{aligned}
    &\frac{1}{2\sqrt{y_{\rm reh}}}\left[ r^{*}\beta_{+}H_0^{(2)}(y_{\rm reh})+r\beta_{-}H_0^{(1)}(y_{\rm reh})\right] \\
    &+\sqrt{y_{\rm reh}}\left[r^{*}\beta_{+}\frac{\text{d}H_0^{(2)}}{\text{d}y}(y_{\rm reh})+r\beta_{-}\frac{\text{d}H_0^{(1)}}{\text{d}y}(y_{\rm reh})\right] \\
    &=i\left(-\gamma_{+}e^{-ik\eta_{\rm reh}}+\gamma_{-}e^{ik\eta_{\rm reh}}\right) \, ,
\end{aligned}
\end{equation}
where
\begin{equation}
    y_{\rm reh}=k\left(\eta_{\rm reh}-\frac{\eta_{\text{kin}}}{2}+\frac{1}{H}\right)\simeq k\eta_{\rm reh} \, ,
\end{equation}
where we have taken into account that $\eta_{\rm reh}\gg \eta_{\rm kin}\gg \eta_{\rm end}$. Summing (subtracting) both expressions gives
\begin{equation}
\begin{aligned}
    &\gamma_{\pm}=\frac{e^{\pm ik\eta_{\rm reh}}}{2}\Biggl\{r^{*}\beta_{+}\Big[H_{0}^{(2)}(y_{\rm reh})\left(\sqrt{y_{\rm reh}}\pm i\frac{1}{2\sqrt{y_{\rm reh}}}\right)\\
    &\mp i \sqrt{y_{\rm reh}}H_{1}^{(2)}(y_{\rm reh})\Big]+r\beta_{-}\Big[H_{0}^{(1)}(y_{\rm reh})\left(\sqrt{y_{\rm reh}}\pm i\frac{1}{2\sqrt{y_{\rm reh}}}\right)\\
    &\mp i \sqrt{y_{\rm reh}}H_{1}^{(1)}(y_{\rm reh})\Big]\Biggl\} \, .
\end{aligned}
\end{equation}
We use Eq. \eqref{asdifubabuhsdf} and take the super-Hubble limit $k\eta_{\rm reh}\ll 1$ to obtain
\begin{equation}
    \gamma_{\pm}=\pm \frac{r\beta_{-}}{2}\frac{i}{\sqrt{y_{\rm reh}}} \, ,
\end{equation}
where $\beta_{-}$ is given by Eq. \eqref{asdifubasdf}. For pure de Sitter, we have the simplified expression
\begin{equation}
    \gamma_{\pm}=\mp \frac{H^2}{2k^2}\sqrt{\frac{\eta_{\rm kin}}{2\eta_{\rm reh}}} \, .
\end{equation}


\bibliography{bibliography}   
\bibliographystyle{spphys}       

%
%

\end{document}